\documentclass[amsmath,trackchanges]{aastex702}
\usepackage{enumitem}

\begin{document}

\title{Quantifying the inside-out formation of disk galaxies at 1.5 $\leq$ z $\leq$ 3.0}

\correspondingauthor{Laura DeGroot}
\email[show]{ldegroot@wooster.edu}  

\author[orcid=0000-0001-9022-665X]{Laura DeGroot}
\affiliation{The College of Wooster, Department of Physics, Wooster, OH, USA}
\email{ldegroot@wooster.edu}  
\email[show]{ldegroot@wooster.edu}

\author[orcid=0000-0002-5269-6527]{Swara Ravindaranath}
\affiliation{Astrophysics Science Division, NASA Goddard Space Flight Center, 8800 Greenbelt Road, Greenbelt, MD 20771, USA} 
\affiliation{Center for Research and Exploration in Space Science \& Technology II, Department of Physics, The Catholic University of America, 620
Michigan Ave., N.E. Washington, DC 20064, USA}
\email{swara.ravindranath@nasa.gov}

\author{Karmellah Buttler}
\affiliation{The College of Wooster, Department of Physics, Wooster, OH, USA}
\email{karmellahbuttler@gmail.com}

\author{Taliah Lansing}
\affiliation{The College of Wooster, Department of Physics, Wooster, OH, USA}
\email{taliah.lansing@gmail.com}

\begin{abstract}

We examine the properties of galaxies at $1.5\le z\le3.0$ in the Extended Groth Strip (EGS) field to explore inside-out disk growth. We take advantage of the high-resolution James Webb Space Telescope (JWST) imaging and imaging from the Hubble Space Telescope (HST) to measure rest-frame optical and rest-frame UV half-light radii of the galaxies at comparable spatial resolution. We also examine the relation between structural properties and star formation rates, mass, and dust attenuation. While previous studies have inferred inside-out disk formation via the comparison of rest-optical and rest-UV radii using HST images, this study advances this knowledge by using the unprecedented resolution of JWST. We find evidence of inside-out disk formation from, on average, larger UV sizes than optical for galaxies across all redshifts. While previous studies suggested that this could be accounted for due to dust, we find that measurements with the improved resolution of the JWST support this size difference, after accounting for dust attenuation. We also observe that majority of galaxies with clumpy morphologies ($n < 0.5$) exhibit larger UV sizes than rest-optical because the star formation is distributed in multiple star forming clumps or mergers. We observe correlations in both the optical and UV for the size-mass relation, which are consistent with previous results but we find lower values for the slope of the power-law relation of the rest-optical fit, based on robust measurements from JWST images compared to the low resolution HST images used in previous studies.

\end{abstract}


\section{Introduction} \label{sec:intro}

During the approximately 3.5 billion years of Cosmic Noon, from $z\sim 3$ to $z\sim 1$, about half of the stellar mass in galaxies observed today was formed. This is believed to have been accompanied by a rapid transformation in galaxy morphologies and significant galaxy growth both before and after the peak in cosmic star-formation around $z\sim 2$ \citep{Bundy_etal_2005, 2011MNRAS.413.2845M, 2020ARA&A..58..661F}. The sizes and morphologies of galaxies are simple probes of the formation and evolution over this period, and when combined with observations over different wavelengths can enable the study of where stars form and therefore how galaxy structure is assembled over time.   

Many past studies have investigated which galaxies are growing over this epoch. Despite the complex nature of their evolution, galaxies have been shown to have a tight and roughly linear correlation between stellar mass and star formation rates \citep{2009ApJ...690..937D, 2012ApJ...754L..29W} and increasing specific star formation rates with redshift, independent of mass \citep{2007ApJ...670..156D, 2011ApJ...730...61K}. Galaxies have also been shown to have relations between star formation rates and structural properties. \citet{2008ApJ...688..770F} found a tight correlation between color and size at a given mass, and when investigating structural parameters of size and S\'ersic index on the SFR-mass diagram,  \citet{2011ApJ...742...96W} found a structural main sequence exists, as seen from a dependence of size and S\'ersic index on the position of galaxies on the known tight correlation between stellar mass and star formation rate (SFR). By including size information to this relation, they observe that these galaxies grow to larger sizes with an increasing zeropoint of the linear SFR-mass relation with lookback time. Other studies have also found tight correlations between the structural properties such as size and stellar mass \citep{2014ApJ...788...28V, 2024ApJ...970..188N}.  The existence of these scaling relations and the prevalence of well-defined disk structure at $z\sim 2$ imply that internal processes shape galaxies and  likely a lower significance of major merger events at this epoch \citep{2020ARA&A..58..661F}. 

Other studies have investigated how these galaxies are growing, including investigating the theory of inside-out disk formation through both H$\alpha$ and stellar continuum sizes and distribution \citep{2012ApJ...747L..28N, 2013ApJ...763L..16N, 2016ApJ...828...27N} or through the comparison of rest-frame UV and optical sizes of galaxies \cite{2024ApJ...970..188N}. In \citet{2012ApJ...747L..28N, 2016ApJ...828...27N}, inside-out disk formation was inferred out to $z \sim 1.5$ through the use of $H\alpha$ maps, finding larger $H\alpha$ sizes compared to stellar continuum. The effects of dust were investigated by \citet{2013ApJ...763L..16N} through $U-V$ color radial profile analysis, which enabled dust corrections and a better understanding of the distribution of the star formation within the galaxies. In the recent study by \cite{2024ApJ...970..188N} using Hubble Space Telescope (HST) imaging, and through comparison with simulations, it was determined that although the most massive galaxies do tend to show steepening of the rest-frame UV stellar mass-size relation in comparison with rest-optical sizes out to $z \sim 3$, this can be attributed completely to the effects of dust attenuation rather than evidence of the inside-out assembly of galaxies.  

Investigating inside-out disk assembly using UV and optical sizes clearly poses challenges with regards to dust. Simulations show that dust attenuation results in the flattening of the rest-UV light distribution due to the fact that dust preferentially attenuates bright sightlines and therefore exhibits large extinction for highly star-forming regions where we would find bright, young stars \citep{2022MNRAS.511.5475M,2020MNRAS.494.5636W}, assuming their models accurately represent star-forming galaxies. This then results in measurements of larger UV sizes due to dust attenuation, and is more prominent for more massive galaxies \citep{2020MNRAS.494.5636W}.

It is now possible to obtain high spatial resolution rest-optical imaging provided by the NIRCam instrument on the James Webb Space Telescope (JWST), which enables a more detailed study of the rest-UV and rest-optical sizes of galaxies from $1.5 \le z \le 3.0$. Through radial profiles and attenuation measurements, dust can be taken into consideration to investigate if star forming galaxies around this epoch show evidence of inside-out disk growth. The layout of this paper is as follows. In sections \ref{sec:data}, we discuss the data for our analysis. In section \ref{sec:analysis}, we discuss the morphological measurement aspects, which is important in both selecting the disk galaxy sample and in obtaining the rest-UV and rest-optical sizes, and we discuss the results from the analysis in section \ref{sec:discuss}. We also assume a flat $\Lambda$CDM cosmology with a Hubble constant of $H_0 = 70$ km s$^{-1}$Mpc$^{-1}$ and cosmological density parameters $\Omega_m = 0.3$ and $\Omega_\Lambda = 0.7$. \\

\section{Data} \label{sec:data}

In this section, we first describe the data set used for the study in Section \ref{subsec: data set}. Next, we describe the analytical methods used to measure the morphological properties of the sample in Section \ref{subsec:morph} followed by a discussion of the final sample selection in Section \ref{subsec:sample}.  

\subsection{Data Set} \label{subsec: data set}

For this study we use rest-frame ultraviolet and rest-frame optical imaging data from the UVCANDELS \citep{2024RNAAS...8...26W}, Cosmic Assembly Near-Infrared Deep Extragalactic Legacy Survey (CANDELS) \citep{2011ApJS..197...35G,2011ApJS..197...36K}, and from the Cosmic Evolution Early Release Science Survey (CEERS) Program, JWST-ERS-1345 \citep{2017jwst.prop.1345F,2023ApJ...946L..13F}. All the {\it HST} and {\it JWST} data used in this paper can be found in MAST: \dataset[10.17909/T94S3X]{http://dx.doi.org/10.17909/T94S3X}, \dataset[10.17909/z7p0-8481]{http://dx.doi.org/10.17909/z7p0-8481}, \dataset[10.17909/8s31-f778]{http://dx.doi.org/10.17909/8s31-f778}. We use 30mas F435W and F606W images from the UVCANDELS and CANDELS surveys, respectively, to achieve rest-UV at 1500 {\AA} over our desired redshift range of $1.5 < z < 3.0$ and the 30mas F150W and F200W NIRCam JWST CEERS \citep{2023ApJ...946L..12B} images to achieve rest-optical at 5500 {\AA}. 

Photometric catalogs are also publicly available on MAST for CANDELS images \citep{2017ApJS..229...32S} and the UVCANDELS images \citep{mehta2024uvcandelscatalogsphotometricredshifts}.  While we refer the reader to the respective publications for the exact details of the photometric techniques used to assemble the multiwavelength catalogs, we briefly summarize the pertinent details here. In brief, for all CANDELS catalogs, the method of flux measurement for the photometric catalog was determined based on the angular resolution of the images. For high-resolution images, photometry was performed with \verb|SExtractor| \citep{1996A&AS..117..393B} in dual-mode with the F160W image and the PSF-matched high-resolution image, the point-spread function (PSF) for each image is convolved to match the F160W resolution, which has the broadest PSF among the HST filters. For low-resolution images, the \verb|TFIT| software \citep{2007PASP..119.1325L} was used to perform the photometry, which uses a morphological template-fitting technique. For the additional photometry included by the UVCANDELS team, F275W and F435W magnitudes are computed using UV-optimized isophotes as described in \cite{2013AJ....146..159T}, which uses isophotes defined on the F606W image to measure fluxes from unconvolved images, rather than PSF-matched images. Adjustment factors are then applied to correct for the aperture and PSF differences to match back to the F160W isophote-based photometry. See  \citet{2024ApJ...972....8S} for additional information about this technique. The photometry for the F150W and F200W CEERS images was measured using the {\it{photutils}} package \citep{bradley_2026_19636730}. Empirical PSFs were derived from each drizzled image using the IRAF DAOPHOT package \citep{1987PASP...99..191S} using bright, non-saturated stars as inputs and then stacking them together.

We also obtain stellar mass \citep{2017ApJS..229...32S}, star formation rate \citep{2019ApJS..243...22B}, and redshift estimates \citep{2023ApJ...942...36K} from the CANDELS collaboration. We use spectroscopic redshifts where available, and supplement with photometric estimates (i.e., we use the \verb|z_best| parameter) in other cases. The normalized median absolute deviation of the differences between the photometric and spectroscopic redshifts in the full CANDELS sample equals 0.0227, and the outlier fraction (defined as the fraction where $|\Delta z/(1 + z)| > 0.15$) equals 0.067 \citep{2023ApJ...942...36K}. Stellar masses used in the study were the median of the six different calculations using exponentially declining SFH, with $\tau$ as a free parameter, and the Chabrier (2003) IMF and taking the nebular lines into account \citep{2017ApJS..229...32S}. To briefly summarize the process of \cite{2019ApJS..243...22B}, SFR estimates were obtained using UV-to-FIR SEDs for the galaxies using a ladder of SFR indicators that were cross-calibrated on relatively massive galaxies with intermediate dust attenuation and low IR fluxes, similar to the method described in \cite{2011ApJ...738..106W}.

 \subsection{Morphology} \label{subsec:morph}

We use the two-dimensional parametric image fitting program GALFIT \citep{2002AJ....124..266P,2010AJ....139.2097P} to model the surface brightness distribution using a S\'ersic function of all galaxies within the desired redshift range in all four observed wavelengths. The S\'ersic function can be expressed in the analytical form 

\begin{equation}
\Sigma(r) = \Sigma_e \exp\left({-\kappa[(r/r_e)^{1/n}-1]}\right),
\end{equation}

where $r_e$ is the effective radius of the galaxy, $\Sigma_e$ is the surface brightness at $r_e$, $n$ is the S\'ersic index, and $\kappa$ is a constant coupled to $n$. This equation reduces to an exponential disk profile for $n=1$, and we could instead replace the effective radius with the scale length, $r_s = 1.678r_e$, the typical radius at which a disk galaxies light profile drops of by a factor of $e$. Each model is convolved with a PSF and optimized using a Levenberg-Marquardt algorithm for $\chi^2$ minimization. Before modeling the galaxies with GALFIT, we first optimally determine image cut out sizes, known as postage stamps, based on galaxy photometric parameters obtained from the Astropy \verb|Photutils| package \citep{larry_bradley_2024_13989456} and using procedures similar to the GALAPAGOS GALFIT automation code \citep{2012MNRAS.422..449B}. Additional objects within the postage stamps were masked out or simultaneously fit with a S\'ersic function depending on the proximity to the central target galaxy. 

Studies have found that a precise background sky level is the most critical systematic in galaxy surface brightness profile fitting \citep{1996A&AS..118..557D,2007ApJS..172..615H}, and since the image cutouts provided for GALFIT are too small for GALFIT to accurately measure the background itself, we make sure to accurately measure it and then make it a fixed parameter. We adopt a flux growth method similar to the process used in GALAPAGOS \citep{2012MNRAS.422..449B} to estimate the local background flux level around each individual object. After masking out other objects in an enlarged postage stamp, we calculate the average flux in elliptical annuli as a function of radius centered around the target galaxy. The background sky level is determined once the slope of the measurements levels off ensuring an accurate and robust background measurement. 

Starting parameters for modeling come from either the photometric catalogs obtained from the Astropy \verb|Photutils| package applied to the CEERS images for rest-optical or from the publicly available CANDELS or UVCANDELS phototometric catalogs for the rest-UV, and all parameters were allowed to vary during the fitting procedure, except for the $x$ and $y$ coordinate for the rest-UV models, which were fixed based on rest-optical GALFIT measured positions. In order to overcome concerns of substantial underestimates of true uncertainties on the model parameters by the fitting process \citep{2007ApJS..172..615H}, we analyze each galaxy ten times with GALFIT, varying all input initial guess parameters by small random variations. Once completed, we average the results to get the final morphological parameters and calculate the uncertainties in the mean as $\delta \bar{x}= \sigma/\sqrt{N}$, where $\sigma$ is the standard deviation and $N$ is the number of completed runs between zero and ten depending on the success of the GALFIT analysis. This uncertainty was added in quadrature to existing averaged GALFIT uncertainties for each parameter. To improve the speed of this process, we run the GALFIT analysis using parallel processing in well-tested Jupyter notebooks. We employ a batch fitting procedure to model each galaxy first in the rest-frame optical wavelength, and then use that determined central position of the galaxy as the fixed central position for the rest-UV model of the galaxy.  \\

\subsection{Final Sample Selection} \label{subsec:sample}

\begin{deluxetable}{l}
\tablecaption{  \label{tab:galaxy_selection_criteria}}
\tablenum{1}

\tablehead{\colhead{Galaxy Selection Criteria}  } 
\startdata
$1.5 \le z \le 3.0$    \\
$8.0 \le \log(M_{\star}/M_{\odot}) \le 11.0$   \\
m$_{\text{F435W}} \le 28.0$, m$_{\text{F606W}} \le 28.5$, m$_{\text{F150W}} \le 29.0$, m$_{\text{F200W}} \le 29.0$   \\
$n_{\text{optical}} \le 2.5$   \\
Flag(GALFIT) = 0$^a$   \\
$(\delta r_e/r_e)_{\text{UV}} < 0.06, (\delta r_e/r_e)_{\text{optical}} < 0.06$  
\enddata
\tablenotetext{a}{Self-defined, $\text{FLAG}> 0$ corresponds to objects with bad or nonexistent GALFIT fit}
\end{deluxetable}

Starting with the available CANDELS and Photutils rest-optical catalogs, we began selecting our galaxy sample by limiting to those with photometric redshift, or spectroscopic if available, to be between $1.5 \le z \le 3.0$. We also require that all galaxies be no fainter than the $5\sigma$ depth for the corresponding rest-optical and rest-UV filter the to ensure that they were bright enough to be modeled accurately ($5\sigma$ depths for each observed band is $m_{\text{AB}} \sim 28$ and $28.5$ for the F435W and F606W HST bands and $29.0$ and $ 29.1$ for F150W and F200W JWST bands, respectively). We additionally limit the sample to galaxies with $ 8.0 \le \log\left(M_*/M_\odot\right) \le 11.0$. We include less massive galaxies than previous studies (i.e. \cite{2014ApJ...788...28V}) because the deep imaging and high resolution of the JWST data allows us to extend our study to lower masses. 

Additionally, previous studies were restricted to high mass galaxies  with $\log\left(M_*/M_\odot\right) \gtrsim 10.0$ in order to ensure mass completeness \citep{2014ApJ...788...28V}. However, in this study, we are focusing on the size comparison between rest-frame UV and optical for which mass-incompleteness is not a factor and enables us to extend our study to lower mass galaxies down to log (M/Msun) = 8.0. The sample was further limited  through a variety of factors once the GALFIT fitting process was complete. Disks were first selected by requiring the measured rest-optical S\'ersic index $n$ to be less than 2.5, $n_{\text{opt}} \le 2.5$. In order to ensure the fitting process was accurate, the resulting rest-optical and rest-UV effective radius errors ($\delta r_{\text{e}}$), FLAG parameters, and reduced chi-squared are used to limit the sample to those with robust fits. We select only galaxies with $\delta r_{\text{e}}/r_{\text{e}} < 0.06$ in both UV and optical. Galaxies were also limited to those with a self-defined FLAG=0 from the GALFIT fitting process indicating that no values were flagged by GALFIT and that the fitting process always converged. Finally, only those with reduced $\chi_{UV}^2 < 2.5$ were selected, as the UV measurements were the most uncertain, and those with low $\chi_{UV}^2$ also had low $\chi_{\text{opt}}^2$. Potential biases could occur due to using a magnitude limited sample over mass limited samples, since magnitude limited samples tend to biased against faint extended sources \citep{2022ApJ...940L..15K, 2023AJ....165...13W} (Kramer et al. 2022; Windhorst et al. 2023). However, our selection of disk galaxies should eliminate these biases, since faint extended sources, such as low surface brightness or ultra diffuse galaxies, tend to be spheroidal or elliptical in structure and would not impact this study \citep{2024ApJ...970..188N,2018RNAAS...2...43C}. This galaxy selection process is outlined in Table \ref{tab:galaxy_selection_criteria}.

\begin{figure}[ht!]
\plotone{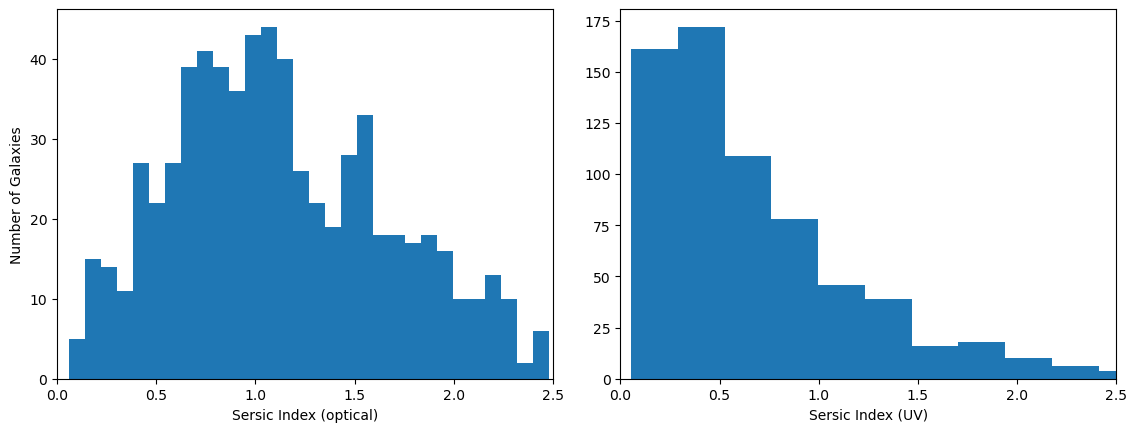}
\caption{The rest-optical and rest-UV S\'ersic index distributions of the final galaxy sample. For the UV sample, we limit the range to $n \le 2.5$ to show the S\'ersic distribution in the disk-like region, but there are 13 galaxies with $n_{\text{UV}} > 2.5$, $n_{\text{optical}} < 2.5$ not shown here. 
\label{fig:galaxy_sample_dist}}
\end{figure}

The final galaxy sample consists of 669 galaxies with 254 at $1.5 \le z < 2.0$, 191 at $2.0 \le z < 2.5$, and 224 at $2.5 \le z \le 3.0$.  We show the distribution of the galaxy S\'ersic indices in rest-optical and rest-UV in Figure \ref{fig:galaxy_sample_dist} and find that the majority of the galaxies identified as disks in the rest-optical are also disks in the rest-UV, with an average optical S\'ersic index $n_{\text{ave,opt}} \approx 1.1$ and an average UV S\'ersic index $n_{\text{ave,UV}} \approx 0.7$. Galaxies with $n<0.5$ usually have clumpy morphologies within an underlying disk or merger morphologies \citep{2006ApJ...652..963R}. We include galaxies with rest-optical $n \le 0.5$, which after visual inspection are primarily disk galaxies with clumpy morphologies.

\section{The UV and Optical Properties of Galaxies} \label{sec:analysis}

For the remainder of the paper, we refer to the effective radius (or half-light radius) of the galaxy as the radius of the galaxy, unless otherwise specified. When discussing optical and UV measurements, we are referring to the rest-optical properties of the galaxy at 5500\r{A} and rest-UV properties at 1500\r{A}. It has been noted that these measurements are less affected by inclination effects than commonly used circularized half-light radii (see also \citet{2014ApJ...788...28V}).

\subsection{Rest-UV and Rest-optical Size Comparison} \label{subsec:sizecomp}

\begin{figure}[ht!]
\includegraphics[width=1.0\textwidth]{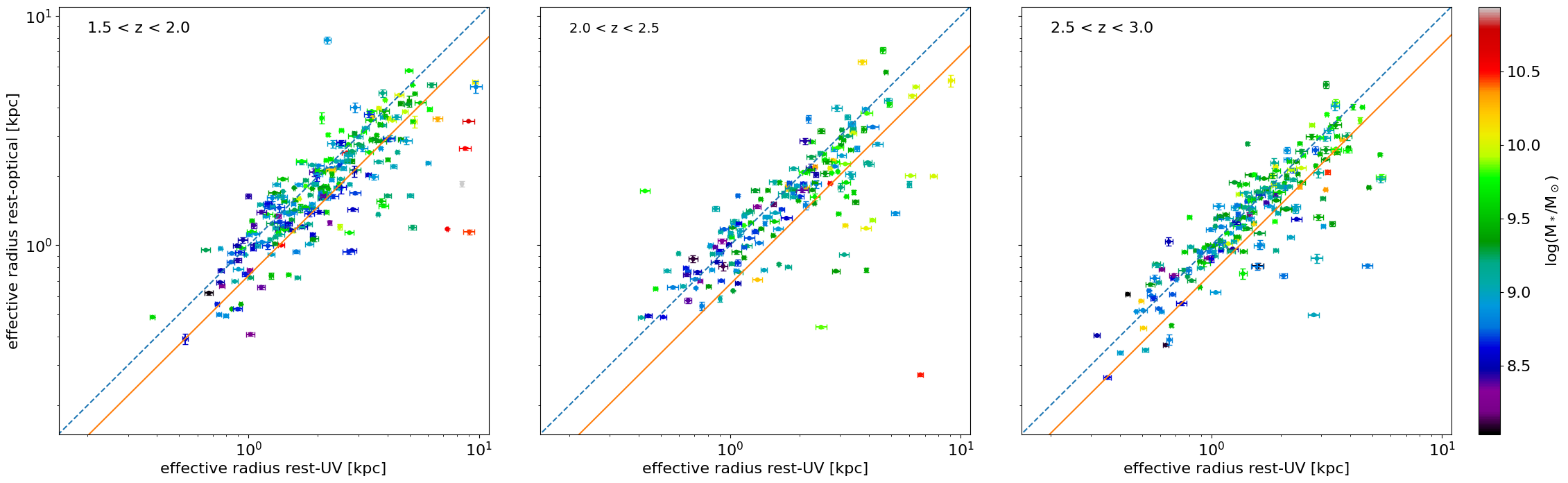}
\caption{The rest-optical effective radius vs. rest-UV effective radius (kpc) separated into $1.5 < z < 2$ on the left,  $2.0 \le z < 2.5$ in the middle, and $ 2.5 \le z \le 3.0$ on the right. The points are colored based on the measured total stellar mass of the galaxy as described by the color bar, and the blue line indicates where the rest-optical and rest-UV radius would be equal for each galaxy.  The orange line is the linear-best fit to the data resulting in slopes of $0.736 \pm 0.02$, $0.68\pm0.03$, and $0.75\pm0.03$, for $1.5\le z < 2.0$, $2.0\le z < 2.5$, and $2.5\le z \le 3.0$, respectively.
\label{fig:sizecomp_mass_3zsplit_spectral}}
\end{figure}

A key aspect of this study is to investigate how the optical and UV sizes compare for galaxies both before and after cosmic noon ($z\sim 2$). If we see larger rest-UV sizes of galaxies compared to optical, this could be interpreted as inside-out disk formation indicating that the galaxies are building up stars on the outskirts of the galaxies beyond the already established population of stars visible in the rest-optical. We examine this in Figure \ref{fig:sizecomp_mass_3zsplit_spectral} and find that for galaxies at $2.5 \le z \le 3.0$, $\sim 62.1\%$ of the galaxies have UV radii that are larger than their measured optical, while at $2.0 \le z < 2.5$, we find that $\sim 64.4\%$ have larger UV radii, and at $1.5 \le z \le 2.0$, $\sim72\%$ have larger UV radii. The average radii of the galaxies grew during this epoch from $\sim1.54$ kpc to $\sim$1.98 kpc in the optical compared to $\sim$1.76 kpc to $\sim$2.48 kpc in the UV.  We note that although there is significant scatter, we find on average larger UV sizes than optical with linear best-fits to the log-sizes with slopes of $0.736 \pm 0.02$, $0.68\pm0.03$, and $0.75\pm0.03$, for $1.5\le z < 2.0$, $2.0\le z < 2.5$, and $2.5\le z \le 3.0$, respectively. Assuming that the UV size traces recent star formation and the optical size traces already established stellar populations, this may be indicative of inside-out disk formation. However, to fully investigate this relation we must consider dust (see section \ref{subsec:dust}).

To further investigate the nature of this observed size difference, we include mass and S\'ersic index to account for the possibility that more massive galaxies with an already well-developed bulge would be more likely to exhibit star formation on their outskirts at their stage of inside-out disk growth, while less massive galaxies likely have more centrally concentrated star formation as they grow from the inside-out. In Figure \ref{fig:sizecomp_mass_3zsplit_spectral}, we have also colored the points by stellar mass to look for trends of inside-out disk growth as a function of mass. For low mass galaxies ($\log_{10}(M_\star/M_\odot) \le 9.0$, most are found along the blue line in the plot indicating equal sizes in both the UV and optical. These low mass galaxies would have a lot of star formation across the entire disk, so the rest-UV and rest-optical light are both dominated by the massive stars with little evidence for inside-out propagation or radial segregation of stellar populations. For higher mass galaxies, we would expect the optical light to be dominated by the bulge, which consists of a more evolved stellar population, and the UV radius would be larger if the inside-out disk growth is the primary mechanism of disk galaxy mass buildup at this stage. In Figure \ref{fig:sizecomp_mass_3zsplit_spectral}, the most massive galaxies are the ones which show larger UV sizes compared to optical sizes.

\begin{figure}[ht!]
\includegraphics[width=1.0\textwidth]{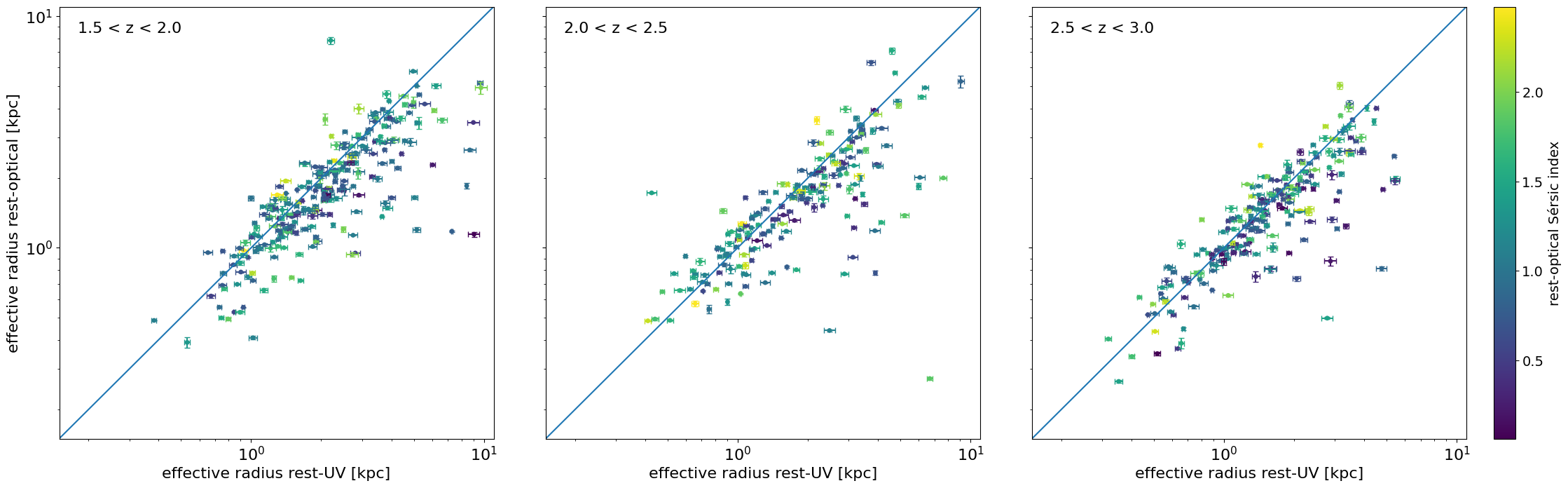}
\caption{The rest-optical effective radius vs. rest-UV effective radius (kpc) separated into $1.5 < z < 2$ on the left,  $2.0 \le z < 2.5$ in the middle, and $ 2.5 \le z \le 3.0$ on the right. The points are colored based on the measured optical S\'ersic index of the galaxy as described by the color bar. The scatter from the solid blue equality line is dominated by clumpy galaxies in the high redshift bin. This is to be expected if the violent instability in gas-rich disks at high redshifts leads to massive clumps \citet{2015ApJ...800...39G,2017MNRAS.464..635M}. 
\label{fig:sizecomp_sersic_3zsplit}}
\end{figure}

\begin{figure}[ht!]
\plotone{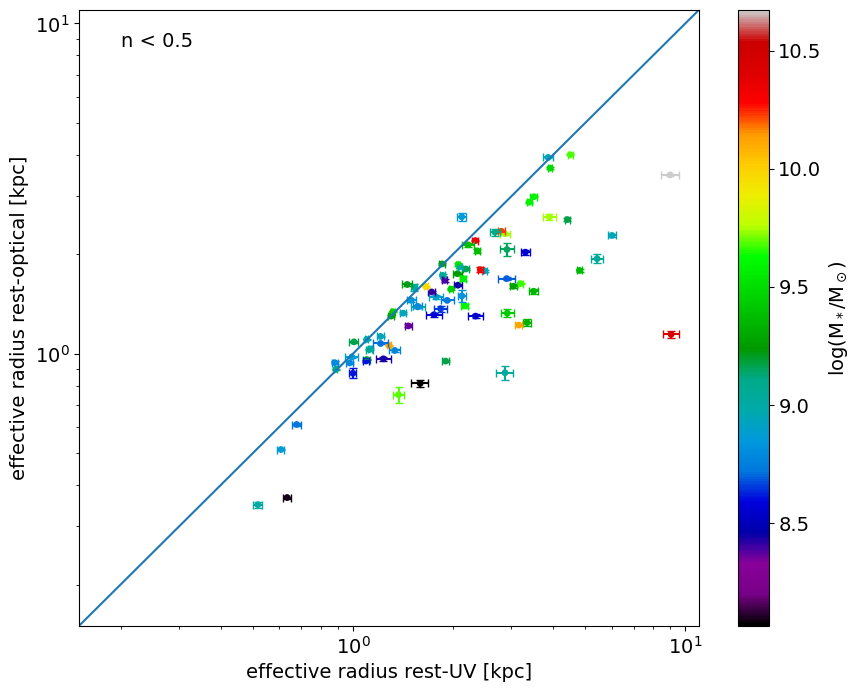}
\caption{The rest-optical effective radius vs. rest-UV effective radius (kpc) for only galaxies with optical S\'ersic index $n < 0.5$. The points are colored based on the estimated stellar mass of the galaxy as described by the color bar.  
\label{fig:sizecomp_nLT05}}
\end{figure}

In order to understand the scatter in the rest-UV vs. rest-optical size relation, we include additional information from the S\'ersic index which relates to the galaxy morphology. Figure \ref{fig:sizecomp_sersic_3zsplit} shows optical vs. UV size with colors indicating the S\'ersic index and separated into three redshift ranges, and Figure \ref{fig:sizecomp_nLT05} has only those with $n \le 0.5$, not separated by redshift, also including mass for all redshifts. In Figure \ref{fig:sizecomp_sersic_3zsplit}, many galaxies with a S\'ersic index of $n\approx 1$, which would be classic disk galaxies, lie along the one-to-one line, indicating that star formation would be occurring throughout the entire galaxy. We find that many galaxies with higher S\'ersic indices, closer to the limits of what would be defined as a disk, lie either close to the one-to-one or even above it, indicating that the well-established population is larger than any recently formed population, so these may be more evolved systems as expected. The galaxies we see to mostly lie below the line in the larger UV-size region are those with $n < 0.5$, which are clumpy galaxies \citep{2006ApJ...652..963R} and are shown separately in Figure \ref{fig:sizecomp_nLT05}, while also colored by mass. Although $n<0.5$ galaxies have clumpy morphologies because of on-going star formation, there is no strong reason to exclude them from an analysis of disk galaxies since previous research indicates that galaxies with $n < 0.5$ often reveal an underlying disk in the rest-optical although the rest-UV morphology is dominated by the UV-bright star forming clumps \citep{2006ApJ...652..963R}.

\begin{figure}[ht!]
\plotone{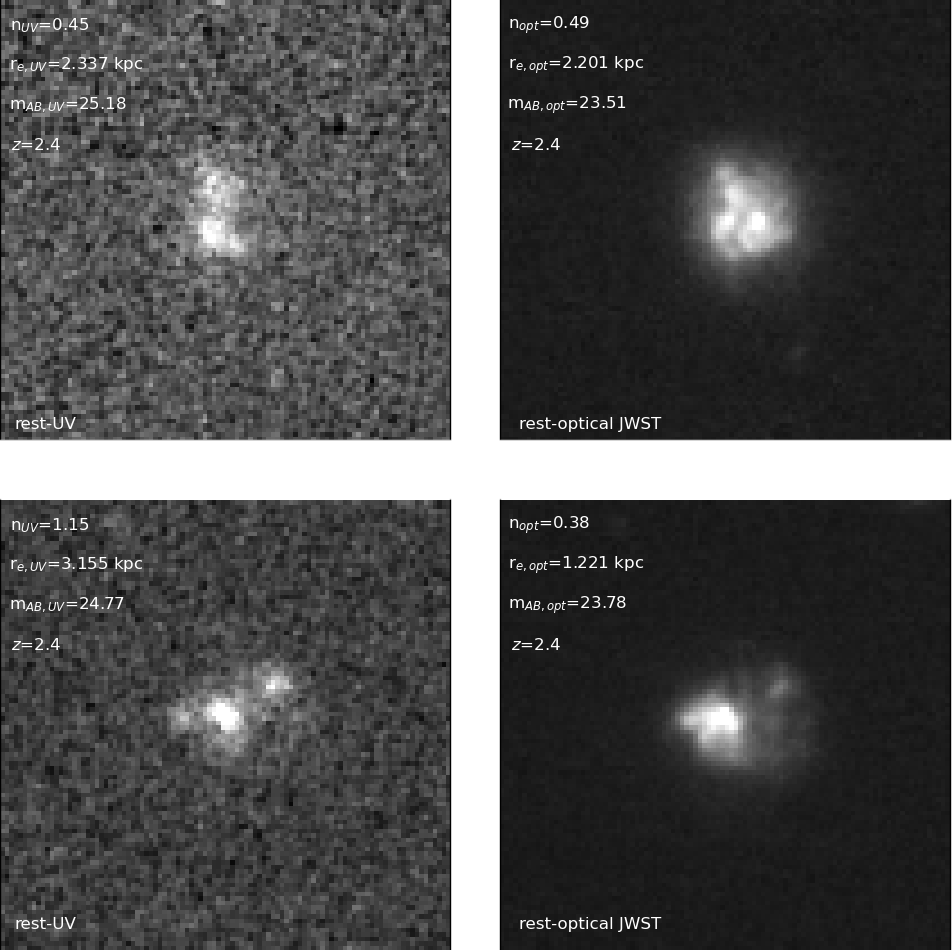}
\caption{Rest-UV and rest-optical postage stamps of two galaxies with S\'ersic index $n < 0.5$. The clumpy morphology of the galaxies is very apparent in both wavelengths. 
\label{fig:clumpy_stamps}}
\end{figure}

Examples of these galaxies are shown in Figure \ref{fig:clumpy_stamps} showing both disk morphology and star-forming clumps within these galaxies. However, due to the star-forming clumps within these galaxies, the UV radii of these galaxies is therefore more extended than the optical suggesting that star formation is occurring more towards the outskirts of these galaxies, which may be stochastic star formation events rather than radial propagation of star formation, thereby mimicking inside-out disk formation, thereby mimicking inside-out disk formation. \\

\subsection{SFR-Mass Relation} \label{subsec:SFRmass}

\begin{figure}[ht!]
\includegraphics[width=1.0\textwidth]{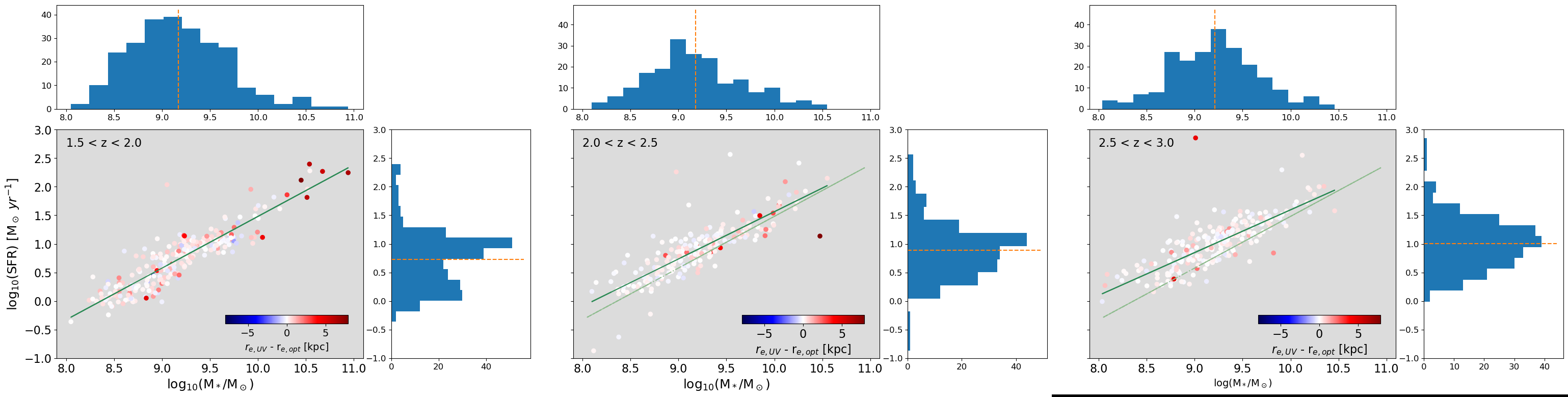}
\caption{The star formation rate vs. stellar mass of the galaxy sample separated at $1.5 < z < 2.0$, $2.0 \le z < 2.5$, and $2.5 \le z \le 3.0$, colored based on the size difference between rest-frame UV and optical. On the right of each plot is a histogram of the $\log_{10}(\text{SFR})$, and above each plot is a histogram of $\log_{10}(M_\star/M_\odot)$. The linear fit to each redshift bin of the galaxy sample is shown in dark green, while the low-redshift ($1.5 \le z < 2.0$) relation is plotted on both of the high redshift bins (dashed light green) to show evolution of the relation with redshift.  The best fit slopes and intercepts, respectively, are $0.90$ and $-7.92$ for $1.5 \le z < 2.0$,  $0.83$ and $-6.70$ for $2.0 \le z < 2.5$, and $0.75$ and $-5.88$ for $2.5 \le z \le 3.0$.
\label{fig:SFR_mass_hist_colorbar}}
\end{figure}

To further interpret these results, we investigated the relation between star formation rate and mass of the galaxy sample at different redshifts, which is shown in Figure \ref{fig:SFR_mass_hist_colorbar}. We observe a linear relation between star formation rate and mass with the zeropoint of the relation increasing with redshift, in agreement with previous research \citep{2017ApJ...839L...5F}. We examine this relation in light of the measured sizes. The color coding on the scatter plot shows white for galaxies with nearly equal optical and UV sizes, while red indicates larger UV sizes and blue indicates larger optical sizes. Figure 6 helps to visualize the larger UV sizes for the massive galaxies in the lowest redshift bin which causes the slope of the relation to be steeper. This could indicate an age gradient with a young stellar population in the disk of massive galaxies which have bulge component that is dominated by an older stellar population. In the highest redshift bin ($2.5<z<3.0$) the massive galaxies have comparable rest-UV and rest-optical sizes from a more uniformly distributed young stellar population.

\subsection{Size-Stellar Mass Relation} \label{subsec:sizemass}

All mass-size relations are derived by fitting the data with $1\sigma$ confidence ranges with

\begin{equation}
R_{\text{eff}} = \text{A}\left(\frac{M_*}{5\times 10^{10}M_\odot}\right)^\alpha,
\end{equation}

where $R_{\text{eff}}$ is the effective radius, $M_*$ is the stellar mass, and $A$ and $\alpha$ are the fit parameters for the log-distribution describing the normalization at $M_*=5\times 10^{10}M_\odot$ and the slope of the relation, respectively. This relation is commonly adopted in literature for star-forming galaxies to characterize the galaxy size distribution as a function of mass \citep{2014ApJ...788...28V, 2024ApJ...970..188N, 2021MNRAS.506..928N, 2019MNRAS.489.4135D, 2024ApJ...964..192I, 2024ApJ...962..176W}. Best-fit parameters are obtained through non-linear least-squares fit optimization using the Levenberg-Marquardt algorithm.

\begin{figure}[ht!]
\includegraphics[width=1.0\textwidth]{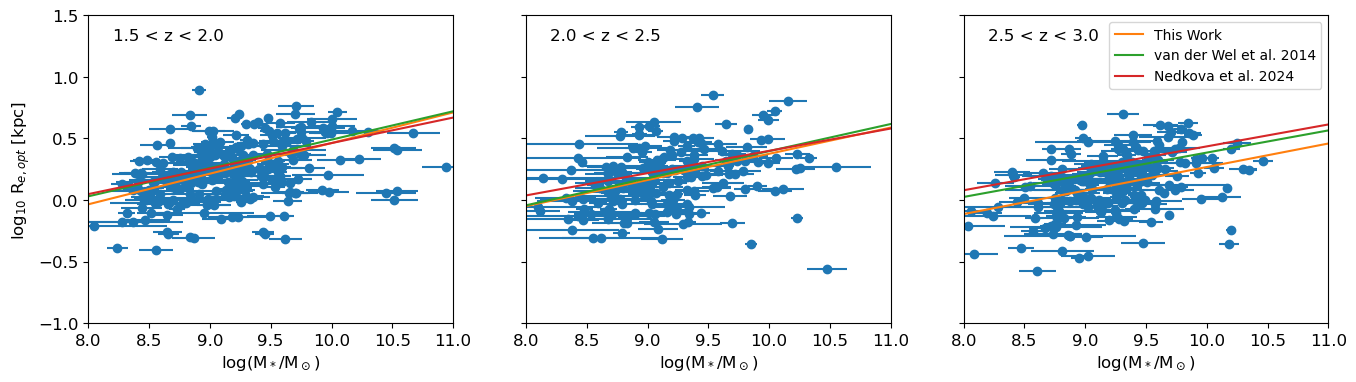}
\caption{The rest-optical effective radius vs. stellar mass divided into $1.5 \le z < 2.0$, $2.0 \le z < 2.5$, and $2.5 \le z \le 3.0$. The best fit lines for the \citet{2014ApJ...788...28V} and \citet{2024ApJ...970..188N} for comparison.
\label{fig:sizemass_optical}}
\end{figure}

\begin{figure}[ht!]
\includegraphics[width=1.0\textwidth]{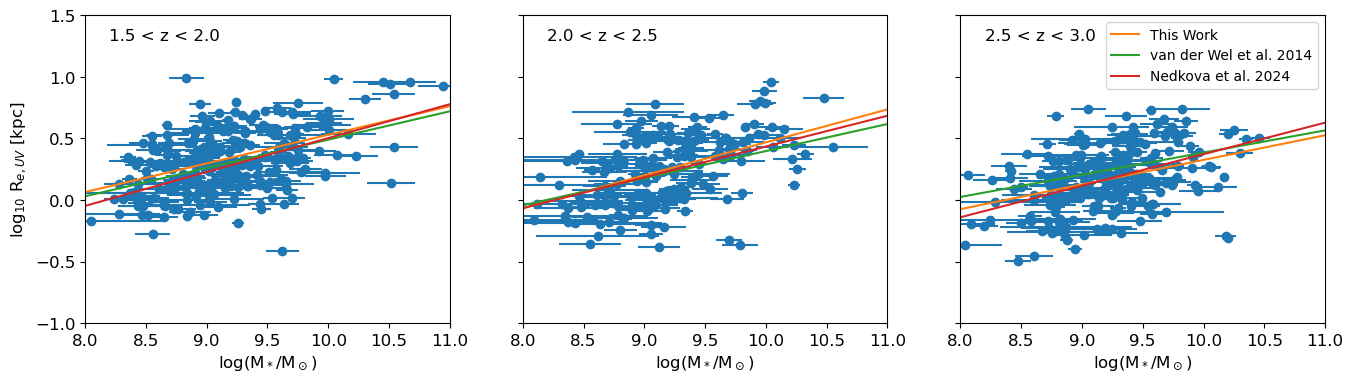}
\caption{The rest-UV effective radius vs. stellar mass divided into $1.5 \le z < 2.0$, $2.0 \le z < 2.5$, and $2.5 \le z \le 3.0$. The best fit line for the \citet{2014ApJ...788...28V} rest-optical relation is included with our and \citet{2024ApJ...970..188N} rest-UV results for comparison. 
\label{fig:sizemass_UV}}
\end{figure}

The size-stellar mass relations for our sample as a function of redshift is shown in Figure \ref{fig:sizemass_optical} for optical sizes and Figure \ref{fig:sizemass_UV} for UV sizes. The best-fit parameters for the optical size-mass relation, which can be directly compared to the results of \citet{2014ApJ...788...28V} and the F160W-selected rest-optical results of \citet{2024ApJ...970..188N}, are shown in Table \ref{tab:sizemass_optical}. We only directly compare our UV size-stellar mass results to that of the UV-selected results \citet{2024ApJ...970..188N} in Table \ref{tab:sizemass_UV}, since \citet{2014ApJ...788...28V}, did not analyze the UV size-mass relation. In general, we find good agreement with previous studies, showing a decreasing exponent of the power law as we go to higher redshift, as well as a decreasing intercept. In the rest-optical relations, the largest disagreement occurs in the highest-redshift bin, $2.5 \le z \le 3.0$, which includes the galaxies in our sample with rest-optical size measured entirely in JWST imaging down to lower masses than previous studies. This could indicate a clear trend trend and perhaps the rather significantly lower slope and intercept for our highest redshift is more indicative of a more robust relation based on the high-resolution measurements. We also find a difference in our UV results, with a slight increase in the power law exponent in the middle redshift range, but \citet{2024ApJ...970..188N} found the exponent to be consistent within the errors, from the lowest to middle redshift range.

\begin{deluxetable*}{lcccccc}
\tabletypesize{\scriptsize}
\tablewidth{0pt} 
\tablenum{2}
\tablecaption{Results from fits to optical size-mass Distribution shown in Figure \ref{fig:sizemass_optical} of the form $R_{\text{eff}}/\text{kpc} = A(M_*/5 \times 10^{10} M_{\odot})^\alpha$ compared to the rest-optical results of \citet{2014ApJ...788...28V} and F160W selected optical relation of \citet{2024ApJ...970..188N}.
\label{tab:sizemass_optical}}
\tablehead{
\colhead{} & \multicolumn{2}{c}{van der Wel} & \multicolumn{2}{c}{Nedkova} &  \multicolumn{2}{c}{This work} \\
\colhead{$z$} & \colhead{$\log$A} & \colhead{$\alpha$} & \colhead{$\log$A} & \colhead{$\alpha$}  & \colhead{$\log$A} & \colhead{$\alpha$} 
} 
\startdata 
$1.75$ & $0.65 \pm 0.01$ & $0.23 \pm 0.01$ & $ 0.606 \pm 0.013 $ & $ 0.207 \pm 0.012 $ & $0.64 \pm 0.06$ & $0.25 \pm 0.03$  \\
$2.25$ & $0.55 \pm 0.01$ & $0.23 \pm 0.01$ & $ 0.525 \pm 0.015 $ & $ 0.181 \pm 0.015 $ & $0.52 \pm 0.04$ & $0.21 \pm 0.03$  \\ 
$2.75$ & $0.51 \pm 0.01$ & $0.18 \pm 0.02$ & $ 0.559 \pm 0.022 $ & $ 0.178 \pm 0.032 $ & $0.40 \pm 0.03$ & $0.19 \pm 0.02$ \\ 
\enddata
\end{deluxetable*}

\begin{deluxetable*}{lcccc}
\tabletypesize{\scriptsize}
\tablewidth{0pt} 
\tablenum{3}
\tablecaption{Results from fits to UV size-mass Distribution shown in Figure \ref{fig:sizemass_UV} of the form $R_{\text{eff}}/\text{kpc} = A(M_*/5 \times 10^{10} M_{\odot})^\alpha$ compared to the rest-optical and rest-UV results of \citet{2014ApJ...788...28V} and \citet{2024ApJ...970..188N}.
\label{tab:sizemass_UV}}
\tablehead{
\colhead{} & \multicolumn{2}{c}{Nedkova} &  \multicolumn{2}{c}{This work} \\
\colhead{$z$} & \colhead{$\log$A} & \colhead{$\alpha$} & \colhead{$\log$A} & \colhead{$\alpha$}  
} 
\startdata 
$1.75$ & $ 0.694 \pm 0.009 $ & $0.275 \pm 0.006 $ & $0.69 \pm 0.06$ & $0.23 \pm 0.04$  \\
$2.25$ & $ 0.608 \pm 0.013 $ & $ 0.250 \pm 0.009 $ & $0.65 \pm 0.07$ & $0.27 \pm 0.05$  \\ 
$2.75$ & $ 0.550 \pm 0.014 $ & $ 0.256 \pm 0.010 $ & $0.46 \pm 0.05$ & $0.20 \pm 0.03$ \\ 
\enddata
\end{deluxetable*}

\subsection{Dust Effects}\label{subsec:dust}

 Dust has been found to impact measured sizes of galaxies in various ways. If the dust is more centrally concentrated, studies have found that dust attenuation results in the flattening of the rest-UV light distribution \citep{2022MNRAS.511.5475M}, and more significantly for more massive galaxies. When the rest-UV size is then measured, such as with GALFIT in this study, this can lead to larger observed UV sizes. If instead the dust was more distributed within the disk of the galaxy, dust attenuation would instead cause sizes to be underestimated. This would imply that a major uncertainty in the interpretation of UV and optical size comparisons is the effect of differential dust extinction, that is, if some parts of the galaxies are more obscured than others \citep{2012ApJ...753..114W, 2013ApJ...763L..16N}.  To asses this, we consider the effect of dust attenuation within the galaxies by using the extinction measured by the CANDELS team \citep{2019ApJS..243...22B}.  In summary, the dust extinction was measured by fitting the observed spectral energy distributions (SEDs) to galaxy templates. The best available SED was used for every galaxy including broad- and medium-band photometry, but not spectroscopy, and extinction along with stellar mass and other physical properties were obtained using the codes \texttt{FAST} \citep{2009ApJ...700..221K, 2018ascl.soft03008K} and \texttt{Synthesizer} \citep{2005ApJ...630...82P,2008ApJ...675..234P}, while the redshift was held fixed to the best redshift estimate, either spectroscopically determined, if possible, or photometric. See \citet{2019ApJS..243...22B} for more information. 

\begin{figure}[ht!]
\includegraphics[width=1.0\textwidth]{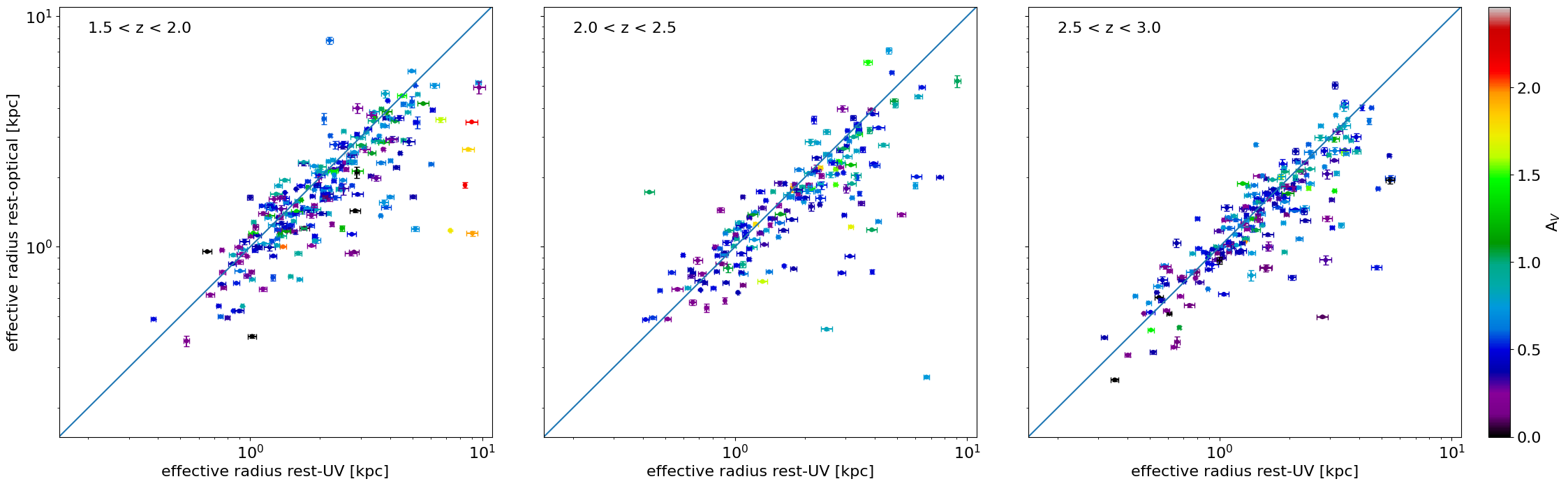}
\caption{Figure 4 is reproduced here, but with the points color-coded by the measured dust attenuation for the galaxies as described by the color bar.  
\label{fig:sizecomp_Av_zsplit}}
\end{figure}

 The rest-optical and rest-UV size comparison is once again investigated across the three different redshift ranges but colored now by dust attenuation as shown in Figure \ref{fig:sizecomp_Av_zsplit}. We find that the highest mass galaxies do have high dust attenuation, indicating that dust could play a significant role in the observed UV sizes and could falsely by implying inside-out disk formation for high mass galaxies. Spatially resolved dust distribution would be required to further study these high-mass  galaxies. However, the majority of galaxies have low dust attenuation, and we observe evidence of on average larger UV sizes compared to optical taking dust into considering.

\begin{figure}[ht!]
\includegraphics[width=1.0\textwidth]{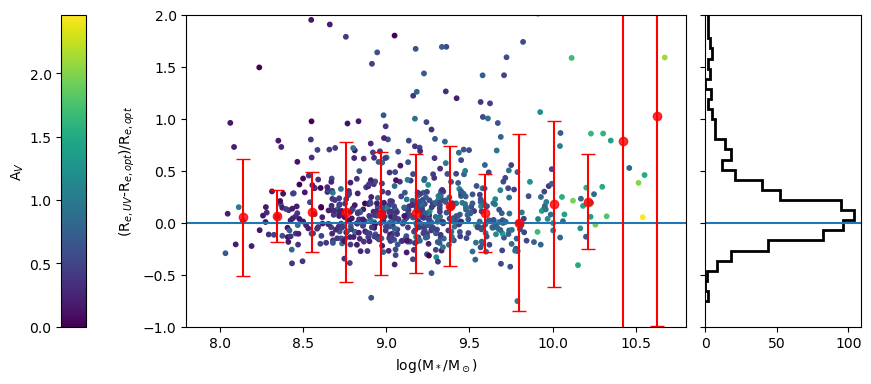}
\caption{The UV-optical size difference of the sample normalized by optical size. Points are colored by attenuation and a running median in included with $1\sigma$ spread as errorbars. The low number of galaxies with high stellar masses makes for very large uncertainties. 
\label{fig:sizecomparison_Av_runningMedian}}
\end{figure}

Additionally, we plot the difference in the rest-UV and optical sizes normalized by the optical size as a function of mass, also colored by dust attenuation as shown in \ref{fig:sizecomparison_Av_runningMedian}. The histogram shows the overall distribution of this result, showing again that more of the sample lies in the larger UV size region of the plot. We also see that the majority of the galaxies are low dust attenuation indicating that dust is not playing a significant role in the results. This also implies inside-out disk formation in the majority of the sample.  We note that the error bars are showing the $1\sigma$ spread in the median data at each of those regions. The spread is large at the high mass region due to very low counts of high mass galaxies remaining in the sample. 

\section{Discussion} \label{sec:discuss}

We have presented rest-UV and rest-optical size comparisons of disk galaxies in the CEERS EGS field in section \ref{subsec:sizecomp}. We find that on average, galaxies are larger in the UV than in the optical. The UV light is representative of younger, short-lived, hot stars, while the rest-optical shows the already established stellar populations. We observe larger UV sizes, which implies that new stellar populations are forming at the outside of the galaxies, on average, indicative of inside-out star formation. This is in agreement with previous literature investigating both $H\alpha$ (instead of UV) and $R$-band sizes of galaxies \citep{2012ApJ...747L..28N,2016ApJ...817L...9N}. 

In the work done by \citet{2024ApJ...970..188N}, through the analysis of the sizes of galaxies observed using HST imaging they measured significant UV sizes compared to optical for only the highest mass galaxies ($\gtrsim 10^{10} M_{\odot}$), but also were found to be highly attenuated. They also compared these results with simulated galaxies from VELA, a set of cosmological zoom-in simulations \citep{2014MNRAS.442.1545C}, and determined that the observed larger UV radii for the high mass galaxies could be completely accounted for due to dust.  We do not observe this same conclusion, as the majority of our sample have low dust attenuation. However, we only have total dust content of the galaxies rather than detailed information about the dust distribution within each galaxy. In future, we would need detailed dust distributions to determine if the dust is centrally concentrated or extended over the disk to fully determine how the distribution of dust impacts our results. While our Figure \ref{fig:sizecomparison_Av_runningMedian} does show larger scatter in the results as indicated by the $1\sigma$ error bars, we do see very low dust attenuation throughout the entire sample all the way to the massive galaxies with $\sim 10^{10} M_{\odot}$. The average sizes larger in the UV than optical including a more significant spread above equality line leaning towards larger UV sizes. 

\begin{figure}[ht!]
\plotone{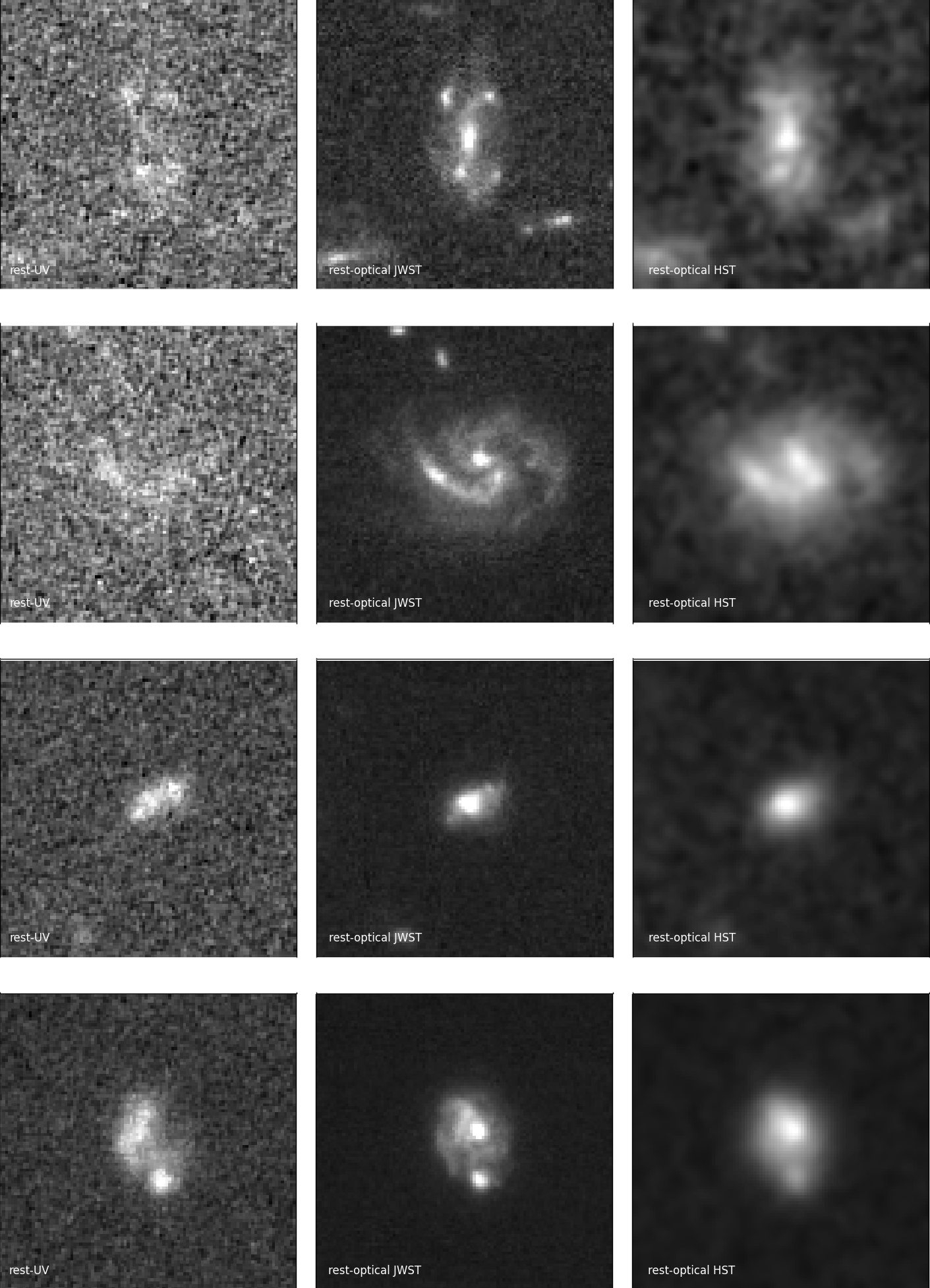}
\caption{The rest-UV (left) and rest-optical JWST (middle) and HST (right) postage stamps of select galaxies.}  
\label{fig:galaxy_stamps_JWST_HST}
\end{figure}

Although our results are based on a smaller sample size than \citet{2024ApJ...970..188N}, by using high-resolution JWST imaging, we are able to better resolve details within galaxies and more robustly model the light distribution of the galaxies. At the observed wavelength of 1500 nm, the HST PSF FWHM is 0.145" while the JWST PSF FWHM is 0.049". Due to the HST resolution, blurring of the galaxies may result in inaccurate estimates of the effective radii (sizes). The average size of a galaxy at $2.5 \le z \le 3.0$ is $\sim 0.22$", so by using the higher resolution of the JWST at this redshift, we are able to obtain robust size measurements due to better pixel sampling of the galaxies based on the significantly smaller PSF FWHM. Figure \ref{fig:galaxy_stamps_JWST_HST} shows a selection of galaxies images in the UV by HST, as well as rest-optical images from both JWST (F150W) and HST (F160W).

We also investigated relations between other structural properties of the galaxy sample. For the stellar mass-size relations in section \ref{subsec:sizemass}, we find that our best-fitting relations for the optical mass-size relation are consistent with \citet{2014ApJ...788...28V} and \citet{2024ApJ...970..188N} and the UV mass-size relations are consistent with \citet{2024ApJ...970..188N} within the uncertainties. This work differs from previous studies in our criteria for selection of disk galaxies and that we include galaxies with stellar masses as low as $10^{8.0} M_{\odot}$, which we discuss in section \ref{subsec:sample} to not lead to any additional bias. Even with this lower mass limit, we confirm agreement between our results and previous studies. Figure \ref{fig:sizemass_UV}, further emphasizes the impact of the smaller PSF FWHM when using JWST. In this figure, only HST data is used to investigate the UV size-mass relation, and we find very consistent results to previous literature since all the studies used the highest resolution data possible from HST. Instead, in Figure \ref{fig:sizemass_optical}, we see a fairly significant difference from the work of \citet{2014ApJ...788...28V} and \citet{2024ApJ...970..188N} at the highest redshift. At this lookback time, we are measuring more robust sizes with JWST data, which better samples the galaxies with many spatially-resolved elements due to the smaller PSF FWHM. Additionally, we extend our study to a higher dynamic range in stellar mass, down to $\log(M_*/M_\odot) \sim 8$ while other studies extend down to only $5 \times 10^9 M_\odot$. By extending the analysis to lower stellar masses, we reduce biases arising from a limited dynamic range in stellar mass.

We examined the SFR-mass relation in section \ref{subsec:SFRmass} and find an increased star formation rate per stellar mass, consistent with previously found correlations \citep{2009ApJ...690..937D}. We also find that the star-forming disk galaxies occupy a tight sequence within the SFR-$M_*$ plane, in agreement with previous studies \citep{2012ApJ...754L..29W, 2011ApJ...738..106W, 2007ApJ...670..156D}, with slopes close to unity across all redshifts, an evolution to shallower slope at higher redshifts, in agreement with \citet{2012ApJ...754L..29W}. While we have fewer high mass galaxies in the high redshift bin, the trend in the slope of the relation as we go to higher redshift indicates that the lower mass galaxies are rapidly forming stars, causing the slope in Figure \ref{fig:SFR_mass_hist_colorbar} to become shallower compared to the slope at the lower redshifts. 

\section{Summary} \label{sec:summary}

In this study, we examine the properties of a sample of 669 disk galaxies at $1.5 \le z \le 3.0$ in the EGS field including their UV and optical sizes, star formation rates, mass, and dust attenuation. While other studies have investigated inside-out disk formation via the comparison of optical and UV radii, our study advances this knowledge by the unprecedented resolution of JWST. While the work of \citet{2024ApJ...970..188N} used a significantly larger sample of galaxies, the study was limited by use of lower resolution HST imaging. By using higher resolution JWST imaging for all rest-optical images, we have been able to better resolve details within galaxies with higher spatial sampling and therefore more accurately modeled the light distribution.  Our main results are highlighted below:

\begin{enumerate}[label=(\roman*)]
        \item We find that, on average, the UV sizes of galaxies across all redshifts are larger than the optical sizes of galaxies. Typically, galaxies with higher mass and lower optical S\'ersic index tend to have larger differences between their UV and optical sizes, overall indicative of inside-out disk formation, with a mild evolution to larger sizes across all masses with redshift, even after accounting for uncertainties.  
        \item Nearly all galaxies with very low optical S\'ersic index ($n_{\text{opt}} < 0.5$), have a larger UV size than optical. These galaxies tend to be disk galaxies with clumpy morphologies. These star-forming clumps tend to extend the UV light distribution outwards, suggesting that star formation is occurring more towards the outskirts of these galaxies, thereby mimicking inside-out disk formation. However, the $n<0.5$ galaxies may also be merging galaxies or stochastic star formation events as opposed to regular star forming disks.

        \item We also used the results from morphology fitting to shed light on some of the well-known parameter relations discussed in the literature. When observing the SFR-stellar mass relation, our results agree with previous result showing increasing star formation rate per stellar mass with increasing redshift from $z = 1.5$ to $z = 3.0$. We also find that our results in both the optical and UV for the size-mass relation are consistent with previous results besides two key differences. We find lower values for the slope and intercept of the power-law relation of the rest-optical fit at the highest redshift range, $2.5 \le z \le 3.0$, which could indicate more robust analysis due to the high resolution of the JWST imaging. We also find slightly higher power-law slope in the middle redshift range $2.0 \le z < 2.5$ for the UV, which is within error of previous results. 
        \item Previous results have found that any difference in UV and optical sizes could be completely accounted for due to dust \citep{2024ApJ...970..188N}. Our investigations have determined that the most massive galaxies in our sample do exhibit high dust attenuation, and therefore could be largely impacted by false large UV sizes due to flattening of the profile as explained by \citet{2024ApJ...970..188N}. However, the less massive galaxies have low dust attenuation indicating that the measured larger UV sizes compared to optical can be considered indicative of inside-out disk growth. 
\end{enumerate} 

This paper has shown that by investigating properties of galaxies in both the rest-optical and rest-UV around cosmic noon, implications for inside-out disk growth are significant as evident by larger UV sizes than optical on average across all redshift bins. We also find agreement with previous studies, even extending to lower stellar masses, when comparing other stellar properties including mass, star formation rate, and size. 

\subsection{Future Work}

When investigating distant galaxies, HST images show that the fraction of clumpy galaxies and disk galaxies with disturbed morphologies increases at $z > 1.5$ \citep{2006ApJ...652..963R}. These clumps are thought to form through gravitational instability in gas-rich turbulent disks \citep{1999ApJ...514...77N, 2004A&A...413..547I, 2004ApJ...611...20I, 2007ApJ...670..237B, 2009ApJ...707L...1B, 2008ApJ...688...67E, 2009ApJ...703..785D, 2010MNRAS.404.2151C, 2014MNRAS.438.1870D, 2016MNRAS.456.2052I}, which has been observationally supported for massive galaxies \citep{2007ApJ...658..763E, 2008A&A...486..741B, 2008ApJ...687...59G, 2011ApJ...733..101G, 2012ApJ...757..120G, 2015ApJ...800...39G, 2016A&A...592A.122H, 2016ApJ...831...78M, 2017ApJ...839L...5F}. It has also been determined that these large clumps seen in distant star-forming galaxies are highly star-forming regions \citep{2015A&A...575A..56B, 2015Natur.521...54Z, 2012ApJ...757..120G, 2012ApJ...753..114W, 2013ApJ...779..135W, 2014ApJ...797..108H, 2016ApJ...831...78M}, which may indicate how galaxy bulges  \citep{2012ApJ...753..114W, 2007ApJ...670..237B, 2008ApJ...688...67E, 2010MNRAS.404.2151C, 2014ApJ...780...57B, 2014MNRAS.443.3675M} and disks \citep{2014MNRAS.442.3206B, 2014MNRAS.441..243I, 2017MNRAS.464.1482S} are evolving with time, although highly debated. With the high spatial resolution of JWST imaging, we can now resolve these detailed star-forming clumps at higher redshift, both before and after cosmic noon, enabling a much more comprehensive study of clump evolution. This future work involving clump evolution in disk galaxies will further this study in helping to understand additional ways in which inside-out disk formation occurs rather than just whether or not it does.

\begin{acknowledgments}

We would like to acknowledge the anonymous referee for the insightful comments.
\software{astropy \citep{2013A&A...558A..33A,2018AJ....156..123A},  
        photutils \citep{larry_bradley_2024_13989456}, GALFIT \citep{2002AJ....124..266P,2010AJ....139.2097P} 
          }

\end{acknowledgments}

\bibliography{InsideOutDisk}{}

@ARTICLE{2018AJ....156..123A,
       author = {{Astropy Collaboration} and {Price-Whelan}, A.~M. and {Sip{\H{o}}cz}, B.~M. and {G{\"u}nther}, H.~M. and {Lim}, P.~L. and {Crawford}, S.~M. and {Conseil}, S. and {Shupe}, D.~L. and {Craig}, M.~W. and {Dencheva}, N. and {Ginsburg}, A. and {VanderPlas}, J.~T. and {Bradley}, L.~D. and {P{\'e}rez-Su{\'a}rez}, D. and {de Val-Borro}, M. and {Aldcroft}, T.~L. and {Cruz}, K.~L. and {Robitaille}, T.~P. and {Tollerud}, E.~J. and {Ardelean}, C. and {Babej}, T. and {Bach}, Y.~P. and {Bachetti}, M. and {Bakanov}, A.~V. and {Bamford}, S.~P. and {Barentsen}, G. and {Barmby}, P. and {Baumbach}, A. and {Berry}, K.~L. and {Biscani}, F. and {Boquien}, M. and {Bostroem}, K.~A. and {Bouma}, L.~G. and {Brammer}, G.~B. and {Bray}, E.~M. and {Breytenbach}, H. and {Buddelmeijer}, H. and {Burke}, D.~J. and {Calderone}, G. and {Cano Rodr{\'\i}guez}, J.~L. and {Cara}, M. and {Cardoso}, J.~V.~M. and {Cheedella}, S. and {Copin}, Y. and {Corrales}, L. and {Crichton}, D. and {D'Avella}, D. and {Deil}, C. and {Depagne}, {\'E}. and {Dietrich}, J.~P. and {Donath}, A. and {Droettboom}, M. and {Earl}, N. and {Erben}, T. and {Fabbro}, S. and {Ferreira}, L.~A. and {Finethy}, T. and {Fox}, R.~T. and {Garrison}, L.~H. and {Gibbons}, S.~L.~J. and {Goldstein}, D.~A. and {Gommers}, R. and {Greco}, J.~P. and {Greenfield}, P. and {Groener}, A.~M. and {Grollier}, F. and {Hagen}, A. and {Hirst}, P. and {Homeier}, D. and {Horton}, A.~J. and {Hosseinzadeh}, G. and {Hu}, L. and {Hunkeler}, J.~S. and {Ivezi{\'c}}, {\v{Z}}. and {Jain}, A. and {Jenness}, T. and {Kanarek}, G. and {Kendrew}, S. and {Kern}, N.~S. and {Kerzendorf}, W.~E. and {Khvalko}, A. and {King}, J. and {Kirkby}, D. and {Kulkarni}, A.~M. and {Kumar}, A. and {Lee}, A. and {Lenz}, D. and {Littlefair}, S.~P. and {Ma}, Z. and {Macleod}, D.~M. and {Mastropietro}, M. and {McCully}, C. and {Montagnac}, S. and {Morris}, B.~M. and {Mueller}, M. and {Mumford}, S.~J. and {Muna}, D. and {Murphy}, N.~A. and {Nelson}, S. and {Nguyen}, G.~H. and {Ninan}, J.~P. and {N{\"o}the}, M. and {Ogaz}, S. and {Oh}, S. and {Parejko}, J.~K. and {Parley}, N. and {Pascual}, S. and {Patil}, R. and {Patil}, A.~A. and {Plunkett}, A.~L. and {Prochaska}, J.~X. and {Rastogi}, T. and {Reddy Janga}, V. and {Sabater}, J. and {Sakurikar}, P. and {Seifert}, M. and {Sherbert}, L.~E. and {Sherwood-Taylor}, H. and {Shih}, A.~Y. and {Sick}, J. and {Silbiger}, M.~T. and {Singanamalla}, S. and {Singer}, L.~P. and {Sladen}, P.~H. and {Sooley}, K.~A. and {Sornarajah}, S. and {Streicher}, O. and {Teuben}, P. and {Thomas}, S.~W. and {Tremblay}, G.~R. and {Turner}, J.~E.~H. and {Terr{\'o}n}, V. and {van Kerkwijk}, M.~H. and {de la Vega}, A. and {Watkins}, L.~L. and {Weaver}, B.~A. and {Whitmore}, J.~B. and {Woillez}, J. and {Zabalza}, V. and {Astropy Contributors}},
        title = "{The Astropy Project: Building an Open-science Project and Status of the v2.0 Core Package}",
      journal = {\aj},
         year = 2018,
        month = sep,
       volume = {156},
       number = {3},
          eid = {123},
        pages = {123},
          doi = {10.3847/1538-3881/aabc4f},
archivePrefix = {arXiv},
       eprint = {1801.02634},
 primaryClass = {astro-ph.IM},
       adsurl = {https://ui.adsabs.harvard.edu/abs/2018AJ....156..123A}
}

@ARTICLE{2013A&A...558A..33A,
       author = {{Astropy Collaboration} and {Robitaille}, Thomas P. and
         {Tollerud}, Erik J. and {Greenfield}, Perry and {Droettboom}, Michael and
         {Bray}, Erik and {Aldcroft}, Tom and {Davis}, Matt and
         {Ginsburg}, Adam and {Price-Whelan}, Adrian M. and
         {Kerzendorf}, Wolfgang E. and {Conley}, Alexander and {Crighton}, Neil and
         {Barbary}, Kyle and {Muna}, Demitri and {Ferguson}, Henry and
         {Grollier}, Fr{\'e}d{\'e}ric and {Parikh}, Madhura M. and
         {Nair}, Prasanth H. and {Unther}, Hans M. and {Deil}, Christoph and
         {Woillez}, Julien and {Conseil}, Simon and {Kramer}, Roban and
         {Turner}, James E.~H. and {Singer}, Leo and {Fox}, Ryan and
         {Weaver}, Benjamin A. and {Zabalza}, Victor and {Edwards}, Zachary I. and
         {Azalee Bostroem}, K. and {Burke}, D.~J. and {Casey}, Andrew R. and
         {Crawford}, Steven M. and {Dencheva}, Nadia and {Ely}, Justin and
         {Jenness}, Tim and {Labrie}, Kathleen and {Lim}, Pey Lian and
         {Pierfederici}, Francesco and {Pontzen}, Andrew and {Ptak}, Andy and
         {Refsdal}, Brian and {Servillat}, Mathieu and {Streicher}, Ole},
        title = "{Astropy: A community Python package for astronomy}",
      journal = {\aap},
         year = "2013",
        month = "Oct",
       volume = {558},
          eid = {A33},
        pages = {A33},
          doi = {10.1051/0004-6361/201322068},
archivePrefix = {arXiv},
       eprint = {1307.6212},
 primaryClass = {astro-ph.IM},
       adsurl = {https://ui.adsabs.harvard.edu/abs/2013A&A...558A..33A}
}

@ARTICLE{2023ApJ...946L..12B,
       author = {{Bagley}, Micaela B. and {Finkelstein}, Steven L. and {Koekemoer}, Anton M. and {Ferguson}, Henry C. and {Arrabal Haro}, Pablo and {Dickinson}, Mark and {Kartaltepe}, Jeyhan S. and {Papovich}, Casey and {P{\'e}rez-Gonz{\'a}lez}, Pablo G. and {Pirzkal}, Nor and {Somerville}, Rachel S. and {Willmer}, Christopher N.~A. and {Yang}, Guang and {Yung}, L.~Y. Aaron and {Fontana}, Adriano and {Grazian}, Andrea and {Grogin}, Norman A. and {Hirschmann}, Michaela and {Kewley}, Lisa J. and {Kirkpatrick}, Allison and {Kocevski}, Dale D. and {Lotz}, Jennifer M. and {Medrano}, Aubrey and {Morales}, Alexa M. and {Pentericci}, Laura and {Ravindranath}, Swara and {Trump}, Jonathan R. and {Wilkins}, Stephen M. and {Calabr{\`o}}, Antonello and {Cooper}, M.~C. and {Costantin}, Luca and {de la Vega}, Alexander and {Hilbert}, Bryan and {Hutchison}, Taylor A. and {Larson}, Rebecca L. and {Lucas}, Ray A. and {McGrath}, Elizabeth J. and {Ryan}, Russell and {Wang}, Xin and {Wuyts}, Stijn},
        title = "{CEERS Epoch 1 NIRCam Imaging: Reduction Methods and Simulations Enabling Early JWST Science Results}",
      journal = {\apjl},
         year = 2023,
        month = mar,
       volume = {946},
       number = {1},
          eid = {L12},
        pages = {L12},
          doi = {10.3847/2041-8213/acbb08},
archivePrefix = {arXiv},
       eprint = {2211.02495},
 primaryClass = {astro-ph.IM},
       adsurl = {https://ui.adsabs.harvard.edu/abs/2023ApJ...946L..12B}
}

@ARTICLE{2012MNRAS.422..449B,
       author = {{Barden}, Marco and {H{\"a}u{\ss}ler}, Boris and {Peng}, Chien Y. and {McIntosh}, Daniel H. and {Guo}, Yicheng},
        title = "{GALAPAGOS: from pixels to parameters}",
      journal = {\mnras},
         year = 2012,
        month = may,
       volume = {422},
       number = {1},
        pages = {449-468},
          doi = {10.1111/j.1365-2966.2012.20619.x},
archivePrefix = {arXiv},
       eprint = {1203.1831},
 primaryClass = {astro-ph.IM},
       adsurl = {https://ui.adsabs.harvard.edu/abs/2012MNRAS.422..449B}
}

@ARTICLE{2019ApJS..243...22B,
       author = {{Barro}, Guillermo and {P{\'e}rez-Gonz{\'a}lez}, Pablo G. and {Cava}, Antonio and {Brammer}, Gabriel and {Pandya}, Viraj and {Eliche Moral}, Carmen and {Esquej}, Pilar and {Dom{\'\i}nguez-S{\'a}nchez}, Helena and {Alcalde Pampliega}, Belen and {Guo}, Yicheng and {Koekemoer}, Anton M. and {Trump}, Jonathan R. and {Ashby}, Matthew L.~N. and {Cardiel}, Nicolas and {Castellano}, Marco and {Conselice}, Christopher J. and {Dickinson}, Mark E. and {Dolch}, Timothy and {Donley}, Jennifer L. and {Espino Briones}, N{\'e}stor and {Faber}, Sandra M. and {Fazio}, Giovanni G. and {Ferguson}, Henry and {Finkelstein}, Steve and {Fontana}, Adriano and {Galametz}, Audrey and {Gardner}, Jonathan P. and {Gawiser}, Eric and {Giavalisco}, Mauro and {Grazian}, Andrea and {Grogin}, Norman A. and {Hathi}, Nimish P. and {Hemmati}, Shoubaneh and {Hern{\'a}n-Caballero}, Antonio and {Kocevski}, Dale and {Koo}, David C. and {Kodra}, Dritan and {Lee}, Kyoung-Soo and {Lin}, Lihwai and {Lucas}, Ray A. and {Mobasher}, Bahram and {McGrath}, Elizabeth J. and {Nandra}, Kirpal and {Nayyeri}, Hooshang and {Newman}, Jeffrey A. and {Pforr}, Janine and {Peth}, Michael and {Rafelski}, Marc and {Rodr{\'\i}guez-Munoz}, Lucia and {Salvato}, Mara and {Stefanon}, Mauro and {van der Wel}, Arjen and {Willner}, Steven P. and {Wiklind}, Tommy and {Wuyts}, Stijn},
        title = "{The CANDELS/SHARDS Multiwavelength Catalog in GOODS-N: Photometry, Photometric Redshifts, Stellar Masses, Emission-line Fluxes, and Star Formation Rates}",
      journal = {\apjs},
         year = 2019,
        month = aug,
       volume = {243},
       number = {2},
          eid = {22},
        pages = {22},
          doi = {10.3847/1538-4365/ab23f2},
archivePrefix = {arXiv},
       eprint = {1908.00569},
 primaryClass = {astro-ph.GA},
       adsurl = {https://ui.adsabs.harvard.edu/abs/2019ApJS..243...22B}
}

@ARTICLE{2014MNRAS.442.3206B,
       author = {{Bassett}, Robert and {Glazebrook}, Karl and {Fisher}, David B. and {Green}, Andrew W. and {Wisnioski}, Emily and {Obreschkow}, Danail and {Cooper}, Erin Mentuch and {Abraham}, Roberto G. and {Damjanov}, Ivana and {McGregor}, Peter J.},
        title = "{DYNAMO - II. Coupled stellar and ionized-gas kinematics in two low-redshift clumpy discs}",
      journal = {\mnras},
         year = 2014,
        month = aug,
       volume = {442},
       number = {4},
        pages = {3206-3221},
          doi = {10.1093/mnras/stu1029},
archivePrefix = {arXiv},
       eprint = {1405.6753},
 primaryClass = {astro-ph.GA},
       adsurl = {https://ui.adsabs.harvard.edu/abs/2014MNRAS.442.3206B}
}

@ARTICLE{1996A&AS..117..393B,
       author = {{Bertin}, E. and {Arnouts}, S.},
        title = "{SExtractor: Software for source extraction.}",
      journal = {\aaps},
         year = 1996,
        month = jun,
       volume = {117},
        pages = {393-404},
          doi = {10.1051/aas:1996164},
       adsurl = {https://ui.adsabs.harvard.edu/abs/1996A&AS..117..393B}
}

@ARTICLE{2007ApJ...670..237B,
       author = {{Bournaud}, Fr{\'e}d{\'e}ric and {Elmegreen}, Bruce G. and {Elmegreen}, Debra Meloy},
        title = "{Rapid Formation of Exponential Disks and Bulges at High Redshift from the Dynamical Evolution of Clump-Cluster and Chain Galaxies}",
      journal = {\apj},
         year = 2007,
        month = nov,
       volume = {670},
       number = {1},
        pages = {237-248},
          doi = {10.1086/522077},
archivePrefix = {arXiv},
       eprint = {0708.0306},
 primaryClass = {astro-ph},
       adsurl = {https://ui.adsabs.harvard.edu/abs/2007ApJ...670..237B}
}

@ARTICLE{2009ApJ...707L...1B,
       author = {{Bournaud}, Fr{\'e}d{\'e}ric and {Elmegreen}, Bruce G. and {Martig}, Marie},
        title = "{The Thick Disks of Spiral Galaxies as Relics from Gas-rich, Turbulent, Clumpy Disks at High Redshift}",
      journal = {\apjl},
         year = 2009,
        month = dec,
       volume = {707},
       number = {1},
        pages = {L1-L5},
          doi = {10.1088/0004-637X/707/1/L1},
archivePrefix = {arXiv},
       eprint = {0910.3677},
 primaryClass = {astro-ph.CO},
       adsurl = {https://ui.adsabs.harvard.edu/abs/2009ApJ...707L...1B}
}

@software{bradley_2026_19636730,
  author       = {Bradley, Larry and
                  Sipőcz, Brigitta M. and
                  Robitaille, T. P. and
                  Tollerud, E. J. and
                  Vinícius, Zé and
                  Deil, Christoph and
                  Barbary, Kyle and
                  Wilson, Tom J. and
                  Busko, Ivo and
                  Donath, Axel and
                  Günther, Hans Moritz and
                  Cara, Mihai and
                  Lim, P. L. and
                  Meßlinger, Sebastian and
                  Conseil, Simon and
                  Droettboom, Michael and
                  Bostroem, K. Azalee and
                  Bray, E. M. and
                  Bratholm, Lars Andersen and
                  Burnett, Zach and
                  Jamieson, William and
                  Ginsburg, Adam and
                  Taranu, Dan and
                  Barentsen, Geert and
                  Craig, Matthew W. and
                  Morris, Brett M. and
                  Perrin, Marshall and
                  Rathi, Shivangee},
  title        = {Photutils},
  month        = apr,
  year         = 2026,
  publisher    = {Zenodo},
  version      = {3.0.0},
  doi          = {10.5281/zenodo.19636730},
  url          = {https://doi.org/10.5281/zenodo.19636730},
  swhid        = {swh:1:dir:5ea90432cd987336a26af201c1b50343b12e2daa
                   ;origin=https://doi.org/10.5281/zenodo.596036;visi
                   t=swh:1:snp:25dd84d95353ec3813baa8785c248017da7192
                   33;anchor=swh:1:rel:253cbde27398f9aea9d7368f386661
                   8c00fe16f9;path=astropy-photutils-3322558
                  },
}

@ARTICLE{2008A&A...486..741B,
       author = {{Bournaud}, F. and {Daddi}, E. and {Elmegreen}, B.~G. and {Elmegreen}, D.~M. and {Nesvadba}, N. and {Vanzella}, E. and {Di Matteo}, P. and {Le Tiran}, L. and {Lehnert}, M. and {Elbaz}, D.},
        title = "{Observations and modeling of a clumpy galaxy at z = 1.6. Spectroscopic clues to the origin and evolution of chain galaxies}",
      journal = {\aap},
         year = 2008,
        month = aug,
       volume = {486},
       number = {3},
        pages = {741-753},
          doi = {10.1051/0004-6361:20079250},
archivePrefix = {arXiv},
       eprint = {0803.3831},
 primaryClass = {astro-ph},
       adsurl = {https://ui.adsabs.harvard.edu/abs/2008A&A...486..741B}
}

@ARTICLE{2015A&A...575A..56B,
       author = {{Bournaud}, F. and {Daddi}, E. and {Wei{\ss}}, A. and {Renaud}, F. and {Mastropietro}, C. and {Teyssier}, R.},
        title = "{Modeling CO emission from hydrodynamic simulations of nearby spirals, starbursting mergers, and high-redshift galaxies}",
      journal = {\aap},
         year = 2015,
        month = mar,
       volume = {575},
          eid = {A56},
        pages = {A56},
          doi = {10.1051/0004-6361/201425078},
archivePrefix = {arXiv},
       eprint = {1409.8157},
 primaryClass = {astro-ph.GA},
       adsurl = {https://ui.adsabs.harvard.edu/abs/2015A&A...575A..56B}
}

@ARTICLE{2014ApJ...780...57B,
       author = {{Bournaud}, Fr{\'e}d{\'e}ric and {Perret}, Valentin and {Renaud}, Florent and {Dekel}, Avishai and {Elmegreen}, Bruce G. and {Elmegreen}, Debra M. and {Teyssier}, Romain and {Amram}, Philippe and {Daddi}, Emanuele and {Duc}, Pierre-Alain and {Elbaz}, David and {Epinat}, Benoit and {Gabor}, Jared M. and {Juneau}, St{\'e}phanie and {Kraljic}, Katarina and {Le Floch'}, Emeric},
        title = "{The Long Lives of Giant Clumps and the Birth of Outflows in Gas-rich Galaxies at High Redshift}",
      journal = {\apj},
         year = 2014,
        month = jan,
       volume = {780},
       number = {1},
          eid = {57},
        pages = {57},
          doi = {10.1088/0004-637X/780/1/57},
archivePrefix = {arXiv},
       eprint = {1307.7136},
 primaryClass = {astro-ph.CO},
       adsurl = {https://ui.adsabs.harvard.edu/abs/2014ApJ...780...57B}
}

@software{larry_bradley_2024_13989456,
  author       = {Larry Bradley and
                  Brigitta Sip{\H o}cz and
                  Thomas Robitaille and
                  Erik Tollerud and
                  Z\`e Vin{\'{\i}}cius and
                  Christoph Deil and
                  Kyle Barbary and
                  Tom J Wilson and
                  Ivo Busko and
                  Axel Donath and
                  Hans Moritz G{\"u}nther and
                  Mihai Cara and
                  P. L. Lim and
                  Sebastian Me{\ss}linger and
                  Simon Conseil and
                  Zach Burnett and
                  Azalee Bostroem and
                  Michael Droettboom and
                  E. M. Bray and
                  Lars Andersen Bratholm and
                  Adam Ginsburg and
                  William Jamieson and
                  Geert Barentsen and
                  Matt Craig and
                  Brett M. Morris and
                  Marshall Perrin and
                  Shivangee Rathi and
                  Sergio Pascual and
                  Iskren Y. Georgiev},
  title        = {astropy/photutils: 2.0.2},
  month        = oct,
  year         = 2024,
  publisher    = {Zenodo},
  version      = {2.0.2},
  doi          = {10.5281/zenodo.13989456},
  url          = {https://doi.org/10.5281/zenodo.13989456},
}

@ARTICLE{Bundy_etal_2005,
       author = {{Bundy}, Kevin and {Ellis}, Richard S. and {Conselice}, Christopher J.},
        title = "{The Mass Assembly Histories of Galaxies of Various Morphologies in the GOODS Fields}",
      journal = {\apj},
         year = 2005,
        month = jun,
       volume = {625},
       number = {2},
        pages = {621-632},
          doi = {10.1086/429549},
archivePrefix = {arXiv},
       eprint = {astro-ph/0502204},
 primaryClass = {astro-ph},
       adsurl = {https://ui.adsabs.harvard.edu/abs/2005ApJ...625..621B}
}

@ARTICLE{2010MNRAS.404.2151C,
       author = {{Ceverino}, Daniel and {Dekel}, Avishai and {Bournaud}, Frederic},
        title = "{High-redshift clumpy discs and bulges in cosmological simulations}",
      journal = {\mnras},
         year = 2010,
        month = jun,
       volume = {404},
       number = {4},
        pages = {2151-2169},
          doi = {10.1111/j.1365-2966.2010.16433.x},
archivePrefix = {arXiv},
       eprint = {0907.3271},
 primaryClass = {astro-ph.CO},
       adsurl = {https://ui.adsabs.harvard.edu/abs/2010MNRAS.404.2151C}
}

@ARTICLE{2014MNRAS.442.1545C,
       author = {{Ceverino}, Daniel and {Klypin}, Anatoly and {Klimek}, Elizabeth S. and {Trujillo-Gomez}, Sebastian and {Churchill}, Christopher W. and {Primack}, Joel and {Dekel}, Avishai},
        title = "{Radiative feedback and the low efficiency of galaxy formation in low-mass haloes at high redshift}",
      journal = {\mnras},
         year = 2014,
        month = aug,
       volume = {442},
       number = {2},
        pages = {1545-1559},
          doi = {10.1093/mnras/stu956},
archivePrefix = {arXiv},
       eprint = {1307.0943},
 primaryClass = {astro-ph.CO},
       adsurl = {https://ui.adsabs.harvard.edu/abs/2014MNRAS.442.1545C}
}

@ARTICLE{2018RNAAS...2...43C,
       author = {{Conselice}, Christopher J.},
        title = "{Ultra-diffuse Galaxies Are a Subset of Cluster Dwarf Elliptical/Spheroidal Galaxies}",
      journal = {Research Notes of the American Astronomical Society},
         year = 2018,
        month = mar,
       volume = {2},
       number = {1},
          eid = {43},
        pages = {43},
          doi = {10.3847/2515-5172/aab7f6},
archivePrefix = {arXiv},
       eprint = {1803.06927},
 primaryClass = {astro-ph.GA},
       adsurl = {https://ui.adsabs.harvard.edu/abs/2018RNAAS...2...43C}
}

@ARTICLE{2007ApJ...670..156D,
       author = {{Daddi}, E. and {Dickinson}, M. and {Morrison}, G. and {Chary}, R. and {Cimatti}, A. and {Elbaz}, D. and {Frayer}, D. and {Renzini}, A. and {Pope}, A. and {Alexander}, D.~M. and {Bauer}, F.~E. and {Giavalisco}, M. and {Huynh}, M. and {Kurk}, J. and {Mignoli}, M.},
        title = "{Multiwavelength Study of Massive Galaxies at z\raisebox{-0.5ex}\textasciitilde2. I. Star Formation and Galaxy Growth}",
      journal = {\apj},
         year = 2007,
        month = nov,
       volume = {670},
       number = {1},
        pages = {156-172},
          doi = {10.1086/521818},
archivePrefix = {arXiv},
       eprint = {0705.2831},
 primaryClass = {astro-ph},
       adsurl = {https://ui.adsabs.harvard.edu/abs/2007ApJ...670..156D}
}

@ARTICLE{2009ApJ...690..937D,
       author = {{Damen}, Maaike and {Labb{\'e}}, Ivo and {Franx}, Marijn and {van Dokkum}, Pieter G. and {Taylor}, Edward N. and {Gawiser}, Eric J.},
        title = "{The Evolution of the Specific Star Formation Rate of Massive Galaxies to z \raisebox{-0.5ex}\textasciitilde 1.8 in the Extended Chandra Deep Field South}",
      journal = {\apj},
         year = 2009,
        month = jan,
       volume = {690},
       number = {1},
        pages = {937-943},
          doi = {10.1088/0004-637X/690/1/937},
archivePrefix = {arXiv},
       eprint = {0809.1426},
 primaryClass = {astro-ph},
       adsurl = {https://ui.adsabs.harvard.edu/abs/2009ApJ...690..937D}
}

@ARTICLE{2009ApJ...703..785D,
       author = {{Dekel}, Avishai and {Sari}, Re'em and {Ceverino}, Daniel},
        title = "{Formation of Massive Galaxies at High Redshift: Cold Streams, Clumpy Disks, and Compact Spheroids}",
      journal = {\apj},
         year = 2009,
        month = sep,
       volume = {703},
       number = {1},
        pages = {785-801},
          doi = {10.1088/0004-637X/703/1/785},
archivePrefix = {arXiv},
       eprint = {0901.2458},
 primaryClass = {astro-ph.GA},
       adsurl = {https://ui.adsabs.harvard.edu/abs/2009ApJ...703..785D}
}

@ARTICLE{2014MNRAS.438.1870D,
       author = {{Dekel}, A. and {Burkert}, A.},
        title = "{Wet disc contraction to galactic blue nuggets and quenching to red nuggets}",
      journal = {\mnras},
         year = 2014,
        month = feb,
       volume = {438},
       number = {2},
        pages = {1870-1879},
          doi = {10.1093/mnras/stt2331},
archivePrefix = {arXiv},
       eprint = {1310.1074},
 primaryClass = {astro-ph.CO},
       adsurl = {https://ui.adsabs.harvard.edu/abs/2014MNRAS.438.1870D}
}

@ARTICLE{1996A&AS..118..557D,
       author = {{de Jong}, R.~S.},
        title = "{Near-infrared and optical broadband surface photometry of 86 face-on disk dominated galaxies. II. A two-dimensional method to determine bulge and disk parameters.}",
      journal = {\aaps},
         year = 1996,
        month = sep,
       volume = {118},
        pages = {557-573},
          doi = {10.48550/arXiv.astro-ph/9601002},
archivePrefix = {arXiv},
       eprint = {astro-ph/9601002},
 primaryClass = {astro-ph},
       adsurl = {https://ui.adsabs.harvard.edu/abs/1996A&AS..118..557D}
}

@ARTICLE{2019MNRAS.489.4135D,
       author = {{Dimauro}, Paola and {Huertas-Company}, Marc and {Daddi}, Emanuele and {P{\'e}rez-Gonz{\'a}lez}, Pablo G. and {Bernardi}, Mariangela and {Caro}, Fernando and {Cattaneo}, Andrea and {H{\"a}u{\ss}ler}, Boris and {Kuchner}, Ulrike and {Shankar}, Francesco and {Barro}, Guillermo and {Buitrago}, Fernando and {Faber}, Sandra M. and {Kocevski}, Dale D. and {Koekemoer}, Anton M. and {Koo}, David C. and {Mei}, Simona and {Peletier}, Reynier and {Primack}, Joel and {Rodriguez-Puebla}, Aldo and {Salvato}, Mara and {Tuccillo}, Diego},
        title = "{The structural properties of classical bulges and discs from z {\ensuremath{\sim}} 2}",
      journal = {\mnras},
         year = 2019,
        month = nov,
       volume = {489},
       number = {3},
        pages = {4135-4154},
          doi = {10.1093/mnras/stz2421},
archivePrefix = {arXiv},
       eprint = {1902.04089},
 primaryClass = {astro-ph.GA},
       adsurl = {https://ui.adsabs.harvard.edu/abs/2019MNRAS.489.4135D}
}

@ARTICLE{2007ApJ...658..763E,
       author = {{Elmegreen}, Debra Meloy and {Elmegreen}, Bruce G. and {Ravindranath}, Swara and {Coe}, Daniel A.},
        title = "{Resolved Galaxies in the Hubble Ultra Deep Field: Star Formation in Disks at High Redshift}",
      journal = {\apj},
         year = 2007,
        month = apr,
       volume = {658},
       number = {2},
        pages = {763-777},
          doi = {10.1086/511667},
archivePrefix = {arXiv},
       eprint = {astro-ph/0701121},
 primaryClass = {astro-ph},
       adsurl = {https://ui.adsabs.harvard.edu/abs/2007ApJ...658..763E}
}

@ARTICLE{2008ApJ...688...67E,
       author = {{Elmegreen}, Bruce G. and {Bournaud}, Fr{\'e}d{\'e}ric and {Elmegreen}, Debra Meloy},
        title = "{Bulge Formation by the Coalescence of Giant Clumps in Primordial Disk Galaxies}",
      journal = {\apj},
         year = 2008,
        month = nov,
       volume = {688},
       number = {1},
        pages = {67-77},
          doi = {10.1086/592190},
archivePrefix = {arXiv},
       eprint = {0808.0716},
 primaryClass = {astro-ph},
       adsurl = {https://ui.adsabs.harvard.edu/abs/2008ApJ...688...67E}
}

@MISC{2017jwst.prop.1345F,
       author = {{Finkelstein}, Steven L. and {Dickinson}, Mark and {Ferguson}, Harry C. and {Grazian}, Andrea and {Grogin}, Norman and {Kartaltepe}, Jeyhan and {Kewley}, Lisa and {Kocevski}, Dale D. and {Koekemoer}, Anton M. and {Lotz}, Jennifer and {Papovich}, Casey and {Pentericci}, Laura and {Perez-Gonzalez}, Pablo G. and {Pirzkal}, Norbert and {Ravindranath}, Swara and {Somerville}, Rachel S. and {Trump}, Jonathan R. and {Wilkins}, Stephen Matthew},
        title = "{The Cosmic Evolution Early Release Science (CEERS) Survey}",
 howpublished = {JWST Proposal ID 1345. Cycle 0 Early Release Science},
         year = 2017,
        month = nov,
        pages = {1345},
       adsurl = {https://ui.adsabs.harvard.edu/abs/2017jwst.prop.1345F}
}

@ARTICLE{2023ApJ...946L..13F,
       author = {{Finkelstein}, Steven L. and {Bagley}, Micaela B. and {Ferguson}, Henry C. and {Wilkins}, Stephen M. and {Kartaltepe}, Jeyhan S. and {Papovich}, Casey and {Yung}, L.~Y. Aaron and {Arrabal Haro}, Pablo and {Behroozi}, Peter and {Dickinson}, Mark and {Kocevski}, Dale D. and {Koekemoer}, Anton M. and {Larson}, Rebecca L. and {Le Bail}, Aur{\'e}lien and {Morales}, Alexa M. and {P{\'e}rez-Gonz{\'a}lez}, Pablo G. and {Burgarella}, Denis and {Dav{\'e}}, Romeel and {Hirschmann}, Michaela and {Somerville}, Rachel S. and {Wuyts}, Stijn and {Bromm}, Volker and {Casey}, Caitlin M. and {Fontana}, Adriano and {Fujimoto}, Seiji and {Gardner}, Jonathan P. and {Giavalisco}, Mauro and {Grazian}, Andrea and {Grogin}, Norman A. and {Hathi}, Nimish P. and {Hutchison}, Taylor A. and {Jha}, Saurabh W. and {Jogee}, Shardha and {Kewley}, Lisa J. and {Kirkpatrick}, Allison and {Long}, Arianna S. and {Lotz}, Jennifer M. and {Pentericci}, Laura and {Pierel}, Justin D.~R. and {Pirzkal}, Nor and {Ravindranath}, Swara and {Ryan}, Russell E. and {Trump}, Jonathan R. and {Yang}, Guang and {Bhatawdekar}, Rachana and {Bisigello}, Laura and {Buat}, V{\'e}ronique and {Calabr{\`o}}, Antonello and {Castellano}, Marco and {Cleri}, Nikko J. and {Cooper}, M.~C. and {Croton}, Darren and {Daddi}, Emanuele and {Dekel}, Avishai and {Elbaz}, David and {Franco}, Maximilien and {Gawiser}, Eric and {Holwerda}, Benne W. and {Huertas-Company}, Marc and {Jaskot}, Anne E. and {Leung}, Gene C.~K. and {Lucas}, Ray A. and {Mobasher}, Bahram and {Pandya}, Viraj and {Tacchella}, Sandro and {Weiner}, Benjamin J. and {Zavala}, Jorge A.},
        title = "{CEERS Key Paper. I. An Early Look into the First 500 Myr of Galaxy Formation with JWST}",
      journal = {\apjl},
         year = 2023,
        month = mar,
       volume = {946},
       number = {1},
          eid = {L13},
        pages = {L13},
          doi = {10.3847/2041-8213/acade4},
archivePrefix = {arXiv},
       eprint = {2211.05792},
 primaryClass = {astro-ph.GA},
       adsurl = {https://ui.adsabs.harvard.edu/abs/2023ApJ...946L..13F}
}

@ARTICLE{2017ApJ...839L...5F,
       author = {{Fisher}, David B. and {Glazebrook}, Karl and {Abraham}, Roberto G. and {Damjanov}, Ivana and {White}, Heidi A. and {Obreschkow}, Danail and {Basset}, Robert and {Bekiaris}, Georgios and {Wisnioski}, Emily and {Green}, Andy and {Bolatto}, Alberto D.},
        title = "{Connecting Clump Sizes in Turbulent Disk Galaxies to Instability Theory}",
      journal = {\apjl},
         year = 2017,
        month = apr,
       volume = {839},
       number = {1},
          eid = {L5},
        pages = {L5},
          doi = {10.3847/2041-8213/aa6478},
archivePrefix = {arXiv},
       eprint = {1703.00458},
 primaryClass = {astro-ph.GA},
       adsurl = {https://ui.adsabs.harvard.edu/abs/2017ApJ...839L...5F}
}

@ARTICLE{2020ARA&A..58..661F,
       author = {{F{\"o}rster Schreiber}, Natascha M. and {Wuyts}, Stijn},
        title = "{Star-Forming Galaxies at Cosmic Noon}",
      journal = {\araa},
         year = 2020,
        month = aug,
       volume = {58},
        pages = {661-725},
          doi = {10.1146/annurev-astro-032620-021910},
archivePrefix = {arXiv},
       eprint = {2010.10171},
 primaryClass = {astro-ph.GA},
       adsurl = {https://ui.adsabs.harvard.edu/abs/2020ARA&A..58..661F}
}

@ARTICLE{2008ApJ...688..770F,
       author = {{Franx}, Marijn and {van Dokkum}, Pieter G. and {F{\"o}rster Schreiber}, Natascha M. and {Wuyts}, Stijn and {Labb{\'e}}, Ivo and {Toft}, Sune},
        title = "{Structure and Star Formation in Galaxies out to z = 3: Evidence for Surface Density Dependent Evolution and Upsizing}",
      journal = {\apj},
         year = 2008,
        month = dec,
       volume = {688},
       number = {2},
        pages = {770-788},
          doi = {10.1086/592431},
archivePrefix = {arXiv},
       eprint = {0808.2642},
 primaryClass = {astro-ph},
       adsurl = {https://ui.adsabs.harvard.edu/abs/2008ApJ...688..770F}
}

@ARTICLE{2008ApJ...687...59G,
       author = {{Genzel}, R. and {Burkert}, A. and {Bouch{\'e}}, N. and {Cresci}, G. and {F{\"o}rster Schreiber}, N.~M. and {Shapley}, A. and {Shapiro}, K. and {Tacconi}, L.~J. and {Buschkamp}, P. and {Cimatti}, A. and {Daddi}, E. and {Davies}, R. and {Eisenhauer}, F. and {Erb}, D.~K. and {Genel}, S. and {Gerhard}, O. and {Hicks}, E. and {Lutz}, D. and {Naab}, T. and {Ott}, T. and {Rabien}, S. and {Renzini}, A. and {Steidel}, C.~C. and {Sternberg}, A. and {Lilly}, S.~J.},
        title = "{From Rings to Bulges: Evidence for Rapid Secular Galaxy Evolution at z \raisebox{-0.5ex}\textasciitilde 2 from Integral Field Spectroscopy in the SINS Survey}",
      journal = {\apj},
         year = 2008,
        month = nov,
       volume = {687},
       number = {1},
        pages = {59-77},
          doi = {10.1086/591840},
archivePrefix = {arXiv},
       eprint = {0807.1184},
 primaryClass = {astro-ph},
       adsurl = {https://ui.adsabs.harvard.edu/abs/2008ApJ...687...59G}
}

@ARTICLE{2011ApJ...733..101G,
       author = {{Genzel}, R. and {Newman}, S. and {Jones}, T. and {F{\"o}rster Schreiber}, N.~M. and {Shapiro}, K. and {Genel}, S. and {Lilly}, S.~J. and {Renzini}, A. and {Tacconi}, L.~J. and {Bouch{\'e}}, N. and {Burkert}, A. and {Cresci}, G. and {Buschkamp}, P. and {Carollo}, C.~M. and {Ceverino}, D. and {Davies}, R. and {Dekel}, A. and {Eisenhauer}, F. and {Hicks}, E. and {Kurk}, J. and {Lutz}, D. and {Mancini}, C. and {Naab}, T. and {Peng}, Y. and {Sternberg}, A. and {Vergani}, D. and {Zamorani}, G.},
        title = "{The Sins Survey of z \raisebox{-0.5ex}\textasciitilde 2 Galaxy Kinematics: Properties of the Giant Star-forming Clumps}",
      journal = {\apj},
         year = 2011,
        month = jun,
       volume = {733},
       number = {2},
          eid = {101},
        pages = {101},
          doi = {10.1088/0004-637X/733/2/101},
archivePrefix = {arXiv},
       eprint = {1011.5360},
 primaryClass = {astro-ph.CO},
       adsurl = {https://ui.adsabs.harvard.edu/abs/2011ApJ...733..101G}
}

@ARTICLE{2011ApJS..197...35G,
       author = {{Grogin}, Norman A. and {Kocevski}, Dale D. and {Faber}, S.~M. and {Ferguson}, Henry C. and {Koekemoer}, Anton M. and {Riess}, Adam G. and {Acquaviva}, Viviana and {Alexander}, David M. and {Almaini}, Omar and {Ashby}, Matthew L.~N. and {Barden}, Marco and {Bell}, Eric F. and {Bournaud}, Fr{\'e}d{\'e}ric and {Brown}, Thomas M. and {Caputi}, Karina I. and {Casertano}, Stefano and {Cassata}, Paolo and {Castellano}, Marco and {Challis}, Peter and {Chary}, Ranga-Ram and {Cheung}, Edmond and {Cirasuolo}, Michele and {Conselice}, Christopher J. and {Roshan Cooray}, Asantha and {Croton}, Darren J. and {Daddi}, Emanuele and {Dahlen}, Tomas and {Dav{\'e}}, Romeel and {de Mello}, Du{\'\i}lia F. and {Dekel}, Avishai and {Dickinson}, Mark and {Dolch}, Timothy and {Donley}, Jennifer L. and {Dunlop}, James S. and {Dutton}, Aaron A. and {Elbaz}, David and {Fazio}, Giovanni G. and {Filippenko}, Alexei V. and {Finkelstein}, Steven L. and {Fontana}, Adriano and {Gardner}, Jonathan P. and {Garnavich}, Peter M. and {Gawiser}, Eric and {Giavalisco}, Mauro and {Grazian}, Andrea and {Guo}, Yicheng and {Hathi}, Nimish P. and {H{\"a}ussler}, Boris and {Hopkins}, Philip F. and {Huang}, Jia-Sheng and {Huang}, Kuang-Han and {Jha}, Saurabh W. and {Kartaltepe}, Jeyhan S. and {Kirshner}, Robert P. and {Koo}, David C. and {Lai}, Kamson and {Lee}, Kyoung-Soo and {Li}, Weidong and {Lotz}, Jennifer M. and {Lucas}, Ray A. and {Madau}, Piero and {McCarthy}, Patrick J. and {McGrath}, Elizabeth J. and {McIntosh}, Daniel H. and {McLure}, Ross J. and {Mobasher}, Bahram and {Moustakas}, Leonidas A. and {Mozena}, Mark and {Nandra}, Kirpal and {Newman}, Jeffrey A. and {Niemi}, Sami-Matias and {Noeske}, Kai G. and {Papovich}, Casey J. and {Pentericci}, Laura and {Pope}, Alexandra and {Primack}, Joel R. and {Rajan}, Abhijith and {Ravindranath}, Swara and {Reddy}, Naveen A. and {Renzini}, Alvio and {Rix}, Hans-Walter and {Robaina}, Aday R. and {Rodney}, Steven A. and {Rosario}, David J. and {Rosati}, Piero and {Salimbeni}, Sara and {Scarlata}, Claudia and {Siana}, Brian and {Simard}, Luc and {Smidt}, Joseph and {Somerville}, Rachel S. and {Spinrad}, Hyron and {Straughn}, Amber N. and {Strolger}, Louis-Gregory and {Telford}, Olivia and {Teplitz}, Harry I. and {Trump}, Jonathan R. and {van der Wel}, Arjen and {Villforth}, Carolin and {Wechsler}, Risa H. and {Weiner}, Benjamin J. and {Wiklind}, Tommy and {Wild}, Vivienne and {Wilson}, Grant and {Wuyts}, Stijn and {Yan}, Hao-Jing and {Yun}, Min S.},
        title = "{CANDELS: The Cosmic Assembly Near-infrared Deep Extragalactic Legacy Survey}",
      journal = {\apjs},
         year = 2011,
        month = dec,
       volume = {197},
       number = {2},
          eid = {35},
        pages = {35},
          doi = {10.1088/0067-0049/197/2/35},
archivePrefix = {arXiv},
       eprint = {1105.3753},
 primaryClass = {astro-ph.CO},
       adsurl = {https://ui.adsabs.harvard.edu/abs/2011ApJS..197...35G}
}

@ARTICLE{2012ApJ...757..120G,
       author = {{Guo}, Yicheng and {Giavalisco}, Mauro and {Ferguson}, Henry C. and {Cassata}, Paolo and {Koekemoer}, Anton M.},
        title = "{Multi-wavelength View of Kiloparsec-scale Clumps in Star-forming Galaxies at z \raisebox{-0.5ex}\textasciitilde 2}",
      journal = {\apj},
         year = 2012,
        month = oct,
       volume = {757},
       number = {2},
          eid = {120},
        pages = {120},
          doi = {10.1088/0004-637X/757/2/120},
archivePrefix = {arXiv},
       eprint = {1110.3800},
 primaryClass = {astro-ph.CO},
       adsurl = {https://ui.adsabs.harvard.edu/abs/2012ApJ...757..120G}
}

@ARTICLE{2015ApJ...800...39G,
       author = {{Guo}, Yicheng and {Ferguson}, Henry C. and {Bell}, Eric F. and {Koo}, David C. and {Conselice}, Christopher J. and {Giavalisco}, Mauro and {Kassin}, Susan and {Lu}, Yu and {Lucas}, Ray and {Mandelker}, Nir and {McIntosh}, Daniel H. and {Primack}, Joel R. and {Ravindranath}, Swara and {Barro}, Guillermo and {Ceverino}, Daniel and {Dekel}, Avishai and {Faber}, Sandra M. and {Fang}, Jerome J. and {Koekemoer}, Anton M. and {Noeske}, Kai and {Rafelski}, Marc and {Straughn}, Amber},
        title = "{Clumpy Galaxies in CANDELS. I. The Definition of UV Clumps and the Fraction of Clumpy Galaxies at 0.5 < z < 3}",
      journal = {\apj},
         year = 2015,
        month = feb,
       volume = {800},
       number = {1},
          eid = {39},
        pages = {39},
          doi = {10.1088/0004-637X/800/1/39},
archivePrefix = {arXiv},
       eprint = {1410.7398},
 primaryClass = {astro-ph.GA},
       adsurl = {https://ui.adsabs.harvard.edu/abs/2015ApJ...800...39G}
}

@ARTICLE{2007ApJS..172..615H,
       author = {{H{\"a}ussler}, Boris and {McIntosh}, Daniel H. and {Barden}, Marco and {Bell}, Eric F. and {Rix}, Hans-Walter and {Borch}, Andrea and {Beckwith}, Steven V.~W. and {Caldwell}, John A.~R. and {Heymans}, Catherine and {Jahnke}, Knud and {Jogee}, Shardha and {Koposov}, Sergey E. and {Meisenheimer}, Klaus and {S{\'a}nchez}, Sebastian F. and {Somerville}, Rachel S. and {Wisotzki}, Lutz and {Wolf}, Christian},
        title = "{GEMS: Galaxy Fitting Catalogs and Testing Parametric Galaxy Fitting Codes: GALFIT and GIM2D}",
      journal = {\apjs},
         year = 2007,
        month = oct,
       volume = {172},
       number = {2},
        pages = {615-633},
          doi = {10.1086/518836},
archivePrefix = {arXiv},
       eprint = {0704.2601},
 primaryClass = {astro-ph},
       adsurl = {https://ui.adsabs.harvard.edu/abs/2007ApJS..172..615H}
}

@ARTICLE{2014ApJ...797..108H,
       author = {{Hemmati}, Shoubaneh and {Miller}, Sarah H. and {Mobasher}, Bahram and {Nayyeri}, Hooshang and {Ferguson}, Henry C. and {Guo}, Yicheng and {Koekemoer}, Anton M. and {Koo}, David C. and {Papovich}, Casey},
        title = "{Kiloparsec-scale Properties of Emission-line Galaxies}",
      journal = {\apj},
         year = 2014,
        month = dec,
       volume = {797},
       number = {2},
          eid = {108},
        pages = {108},
          doi = {10.1088/0004-637X/797/2/108},
archivePrefix = {arXiv},
       eprint = {1409.4791},
 primaryClass = {astro-ph.GA},
       adsurl = {https://ui.adsabs.harvard.edu/abs/2014ApJ...797..108H}
}

@ARTICLE{2016A&A...592A.122H,
       author = {{Hinojosa-Go{\~n}i}, R. and {Mu{\~n}oz-Tu{\~n}{\'o}n}, C. and {M{\'e}ndez-Abreu}, J.},
        title = "{Starburst galaxies in the COSMOS field: clumpy star-formation at redshift 0 < z < 0.5}",
      journal = {\aap},
         year = 2016,
        month = aug,
       volume = {592},
          eid = {A122},
        pages = {A122},
          doi = {10.1051/0004-6361/201527066},
archivePrefix = {arXiv},
       eprint = {1604.01698},
 primaryClass = {astro-ph.GA},
       adsurl = {https://ui.adsabs.harvard.edu/abs/2016A&A...592A.122H}
}

@ARTICLE{2004ApJ...611...20I,
       author = {{Immeli}, Andreas and {Samland}, Markus and {Westera}, Pieter and {Gerhard}, Ortwin},
        title = "{Subgalactic Clumps at High Redshift: A Fragmentation Origin?}",
      journal = {\apj},
         year = 2004,
        month = aug,
       volume = {611},
       number = {1},
        pages = {20-25},
          doi = {10.1086/422179},
archivePrefix = {arXiv},
       eprint = {astro-ph/0406135},
 primaryClass = {astro-ph},
       adsurl = {https://ui.adsabs.harvard.edu/abs/2004ApJ...611...20I}
}

@ARTICLE{2004A&A...413..547I,
       author = {{Immeli}, A. and {Samland}, M. and {Gerhard}, O. and {Westera}, P.},
        title = "{Gas physics, disk fragmentation,  and bulge formation in young galaxies}",
      journal = {\aap},
         year = 2004,
        month = jan,
       volume = {413},
        pages = {547-561},
          doi = {10.1051/0004-6361:20034282},
archivePrefix = {arXiv},
       eprint = {astro-ph/0312139},
 primaryClass = {astro-ph},
       adsurl = {https://ui.adsabs.harvard.edu/abs/2004A&A...413..547I}
}

@ARTICLE{2014MNRAS.441..243I,
       author = {{Inoue}, Shigeki and {Saitoh}, Takayuki R.},
        title = "{Properties of thick discs formed in clumpy galaxies}",
      journal = {\mnras},
         year = 2014,
        month = jun,
       volume = {441},
       number = {1},
        pages = {243-255},
          doi = {10.1093/mnras/stu544},
archivePrefix = {arXiv},
       eprint = {1402.5986},
 primaryClass = {astro-ph.GA},
       adsurl = {https://ui.adsabs.harvard.edu/abs/2014MNRAS.441..243I}
}

@ARTICLE{2016MNRAS.456.2052I,
       author = {{Inoue}, Shigeki and {Dekel}, Avishai and {Mandelker}, Nir and {Ceverino}, Daniel and {Bournaud}, Fr{\'e}d{\'e}ric and {Primack}, Joel},
        title = "{Non-linear violent disc instability with high Toomre's Q in high-redshift clumpy disc galaxies}",
      journal = {\mnras},
         year = 2016,
        month = feb,
       volume = {456},
       number = {2},
        pages = {2052-2069},
          doi = {10.1093/mnras/stv2793},
archivePrefix = {arXiv},
       eprint = {1510.07695},
 primaryClass = {astro-ph.GA},
       adsurl = {https://ui.adsabs.harvard.edu/abs/2016MNRAS.456.2052I}
}

@ARTICLE{2024ApJ...964..192I,
       author = {{Ito}, Kei and {Valentino}, Francesco and {Brammer}, Gabriel and {Faisst}, Andreas L. and {Gillman}, Steven and {G{\'o}mez-Guijarro}, Carlos and {Gould}, Katriona M.~L. and {Heintz}, Kasper E. and {Ilbert}, Olivier and {Jespersen}, Christian Kragh and {Kokorev}, Vasily and {Kubo}, Mariko and {Magdis}, Georgios E. and {McPartland}, Conor J.~R. and {Onodera}, Masato and {Rizzo}, Francesca and {Tanaka}, Masayuki and {Toft}, Sune and {Vijayan}, Aswin P. and {Weaver}, John R. and {Whitaker}, Katherine E. and {Wright}, Lillian},
        title = "{Size{\textendash}Stellar Mass Relation and Morphology of Quiescent Galaxies at z {\ensuremath{\geq}} 3 in Public JWST Fields}",
      journal = {\apj},
         year = 2024,
        month = apr,
       volume = {964},
       number = {2},
          eid = {192},
        pages = {192},
          doi = {10.3847/1538-4357/ad2512},
archivePrefix = {arXiv},
       eprint = {2307.06994},
 primaryClass = {astro-ph.GA},
       adsurl = {https://ui.adsabs.harvard.edu/abs/2024ApJ...964..192I}
}

@ARTICLE{2011ApJ...730...61K,
       author = {{Karim}, A. and {Schinnerer}, E. and {Mart{\'\i}nez-Sansigre}, A. and {Sargent}, M.~T. and {van der Wel}, A. and {Rix}, H. -W. and {Ilbert}, O. and {Smol{\v{c}}i{\'c}}, V. and {Carilli}, C. and {Pannella}, M. and {Koekemoer}, A.~M. and {Bell}, E.~F. and {Salvato}, M.},
        title = "{The Star Formation History of Mass-selected Galaxies in the COSMOS Field}",
      journal = {\apj},
         year = 2011,
        month = apr,
       volume = {730},
       number = {2},
          eid = {61},
        pages = {61},
          doi = {10.1088/0004-637X/730/2/61},
archivePrefix = {arXiv},
       eprint = {1011.6370},
 primaryClass = {astro-ph.CO},
       adsurl = {https://ui.adsabs.harvard.edu/abs/2011ApJ...730...61K}
}

@ARTICLE{2023ApJ...942...36K,
       author = {{Kodra}, Dritan and {Andrews}, Brett H. and {Newman}, Jeffrey A. and {Finkelstein}, Steven L. and {Fontana}, Adriano and {Hathi}, Nimish and {Salvato}, Mara and {Wiklind}, Tommy and {Wuyts}, Stijn and {Broussard}, Adam and {Chartab}, Nima and {Conselice}, Christopher and {Cooper}, M.~C. and {Dekel}, Avishai and {Dickinson}, Mark and {Ferguson}, Henry C. and {Gawiser}, Eric and {Grogin}, Norman A. and {Iyer}, Kartheik and {Kartaltepe}, Jeyhan and {Kassin}, Susan and {Koekemoer}, Anton M. and {Koo}, David C. and {Lucas}, Ray A. and {Mantha}, Kameswara Bharadwaj and {McIntosh}, Daniel H. and {Mobasher}, Bahram and {Pacifici}, Camilla and {P{\'e}rez-Gonz{\'a}lez}, Pablo G. and {Santini}, Paola},
        title = "{Optimized Photometric Redshifts for the Cosmic Assembly Near-infrared Deep Extragalactic Legacy Survey (CANDELS)}",
      journal = {\apj},
         year = 2023,
        month = jan,
       volume = {942},
       number = {1},
          eid = {36},
        pages = {36},
          doi = {10.3847/1538-4357/ac9f12},
archivePrefix = {arXiv},
       eprint = {2210.01140},
 primaryClass = {astro-ph.GA},
       adsurl = {https://ui.adsabs.harvard.edu/abs/2023ApJ...942...36K}
}

@ARTICLE{2011ApJS..197...36K,
       author = {{Koekemoer}, Anton M. and {Faber}, S.~M. and {Ferguson}, Henry C. and {Grogin}, Norman A. and {Kocevski}, Dale D. and {Koo}, David C. and {Lai}, Kamson and {Lotz}, Jennifer M. and {Lucas}, Ray A. and {McGrath}, Elizabeth J. and {Ogaz}, Sara and {Rajan}, Abhijith and {Riess}, Adam G. and {Rodney}, Steve A. and {Strolger}, Louis and {Casertano}, Stefano and {Castellano}, Marco and {Dahlen}, Tomas and {Dickinson}, Mark and {Dolch}, Timothy and {Fontana}, Adriano and {Giavalisco}, Mauro and {Grazian}, Andrea and {Guo}, Yicheng and {Hathi}, Nimish P. and {Huang}, Kuang-Han and {van der Wel}, Arjen and {Yan}, Hao-Jing and {Acquaviva}, Viviana and {Alexander}, David M. and {Almaini}, Omar and {Ashby}, Matthew L.~N. and {Barden}, Marco and {Bell}, Eric F. and {Bournaud}, Fr{\'e}d{\'e}ric and {Brown}, Thomas M. and {Caputi}, Karina I. and {Cassata}, Paolo and {Challis}, Peter J. and {Chary}, Ranga-Ram and {Cheung}, Edmond and {Cirasuolo}, Michele and {Conselice}, Christopher J. and {Roshan Cooray}, Asantha and {Croton}, Darren J. and {Daddi}, Emanuele and {Dav{\'e}}, Romeel and {de Mello}, Duilia F. and {de Ravel}, Loic and {Dekel}, Avishai and {Donley}, Jennifer L. and {Dunlop}, James S. and {Dutton}, Aaron A. and {Elbaz}, David and {Fazio}, Giovanni G. and {Filippenko}, Alexei V. and {Finkelstein}, Steven L. and {Frazer}, Chris and {Gardner}, Jonathan P. and {Garnavich}, Peter M. and {Gawiser}, Eric and {Gruetzbauch}, Ruth and {Hartley}, Will G. and {H{\"a}ussler}, Boris and {Herrington}, Jessica and {Hopkins}, Philip F. and {Huang}, Jia-Sheng and {Jha}, Saurabh W. and {Johnson}, Andrew and {Kartaltepe}, Jeyhan S. and {Khostovan}, Ali A. and {Kirshner}, Robert P. and {Lani}, Caterina and {Lee}, Kyoung-Soo and {Li}, Weidong and {Madau}, Piero and {McCarthy}, Patrick J. and {McIntosh}, Daniel H. and {McLure}, Ross J. and {McPartland}, Conor and {Mobasher}, Bahram and {Moreira}, Heidi and {Mortlock}, Alice and {Moustakas}, Leonidas A. and {Mozena}, Mark and {Nandra}, Kirpal and {Newman}, Jeffrey A. and {Nielsen}, Jennifer L. and {Niemi}, Sami and {Noeske}, Kai G. and {Papovich}, Casey J. and {Pentericci}, Laura and {Pope}, Alexandra and {Primack}, Joel R. and {Ravindranath}, Swara and {Reddy}, Naveen A. and {Renzini}, Alvio and {Rix}, Hans-Walter and {Robaina}, Aday R. and {Rosario}, David J. and {Rosati}, Piero and {Salimbeni}, Sara and {Scarlata}, Claudia and {Siana}, Brian and {Simard}, Luc and {Smidt}, Joseph and {Snyder}, Diana and {Somerville}, Rachel S. and {Spinrad}, Hyron and {Straughn}, Amber N. and {Telford}, Olivia and {Teplitz}, Harry I. and {Trump}, Jonathan R. and {Vargas}, Carlos and {Villforth}, Carolin and {Wagner}, Cory R. and {Wandro}, Pat and {Wechsler}, Risa H. and {Weiner}, Benjamin J. and {Wiklind}, Tommy and {Wild}, Vivienne and {Wilson}, Grant and {Wuyts}, Stijn and {Yun}, Min S.},
        title = "{CANDELS: The Cosmic Assembly Near-infrared Deep Extragalactic Legacy Survey{\textemdash}The Hubble Space Telescope Observations, Imaging Data Products, and Mosaics}",
      journal = {\apjs},
         year = 2011,
        month = dec,
       volume = {197},
       number = {2},
          eid = {36},
        pages = {36},
          doi = {10.1088/0067-0049/197/2/36},
archivePrefix = {arXiv},
       eprint = {1105.3754},
 primaryClass = {astro-ph.CO},
       adsurl = {https://ui.adsabs.harvard.edu/abs/2011ApJS..197...36K}
}

@ARTICLE{2022ApJ...940L..15K,
       author = {{Kramer}, Darby M. and {Carleton}, Timothy and {Cohen}, Seth. H. and {Jansen}, Rolf and {Windhorst}, Rogier A. and {Grogin}, Norman and {Koekemoer}, Anton and {MacKenty}, John W. and {Pirzkal}, Nor},
        title = "{SKYSURF-3: Testing Crowded Object Catalogs in the Hubble eXtreme Deep Field Mosaics to Study Sample Incompleteness from an Extragalactic Background Light Perspective}",
      journal = {\apjl},
         year = 2022,
        month = nov,
       volume = {940},
       number = {1},
          eid = {L15},
        pages = {L15},
          doi = {10.3847/2041-8213/ac9cca},
archivePrefix = {arXiv},
       eprint = {2208.07218},
 primaryClass = {astro-ph.CO},
       adsurl = {https://ui.adsabs.harvard.edu/abs/2022ApJ...940L..15K}
}

@ARTICLE{2009ApJ...700..221K,
       author = {{Kriek}, Mariska and {van Dokkum}, Pieter G. and {Labb{\'e}}, Ivo and {Franx}, Marijn and {Illingworth}, Garth D. and {Marchesini}, Danilo and {Quadri}, Ryan F.},
        title = "{An Ultra-Deep Near-Infrared Spectrum of a Compact Quiescent Galaxy at z = 2.2}",
      journal = {\apj},
         year = 2009,
        month = jul,
       volume = {700},
       number = {1},
        pages = {221-231},
          doi = {10.1088/0004-637X/700/1/221},
archivePrefix = {arXiv},
       eprint = {0905.1692},
 primaryClass = {astro-ph.CO},
       adsurl = {https://ui.adsabs.harvard.edu/abs/2009ApJ...700..221K}
}

@software{2018ascl.soft03008K,
       author = {{Kriek}, Mariska and {van Dokkum}, Pieter G. and {Labb{\'e}}, Ivo and {Franx}, Marijn and {Illingworth}, Garth D. and {Marchesini}, Danilo and {Quadri}, Ryan F. and {Aird}, James and {Coil}, Alison L. and {Georgakakis}, Antonis},
        title = "{FAST: Fitting and Assessment of Synthetic Templates}",
 howpublished = {Astrophysics Source Code Library, record ascl:1803.008},
         year = 2018,
        month = mar,
          eid = {ascl:1803.008},
archivePrefix = {ascl},
       eprint = {1803.008},
       adsurl = {https://ui.adsabs.harvard.edu/abs/2018ascl.soft03008K}
}

@ARTICLE{2007PASP..119.1325L,
       author = {{Laidler}, Victoria G. and {Papovich}, Casey and {Grogin}, Norman A. and {Idzi}, Rafal and {Dickinson}, Mark and {Ferguson}, Henry C. and {Hilbert}, Bryan and {Clubb}, Kelsey and {Ravindranath}, Swara},
        title = "{TFIT: A Photometry Package Using Prior Information for Mixed-Resolution Data Sets}",
      journal = {\pasp},
         year = 2007,
        month = nov,
       volume = {119},
       number = {861},
        pages = {1325-1344},
          doi = {10.1086/523898},
       adsurl = {https://ui.adsabs.harvard.edu/abs/2007PASP..119.1325L}
}

@ARTICLE{2014MNRAS.443.3675M,
       author = {{Mandelker}, Nir and {Dekel}, Avishai and {Ceverino}, Daniel and {Tweed}, Dylan and {Moody}, Christopher E. and {Primack}, Joel},
        title = "{The population of giant clumps in simulated high-z galaxies: in situ and ex situ migration and survival}",
      journal = {\mnras},
         year = 2014,
        month = oct,
       volume = {443},
       number = {4},
        pages = {3675-3702},
          doi = {10.1093/mnras/stu1340},
archivePrefix = {arXiv},
       eprint = {1311.0013},
 primaryClass = {astro-ph.CO},
       adsurl = {https://ui.adsabs.harvard.edu/abs/2014MNRAS.443.3675M}
}

@ARTICLE{2017MNRAS.464..635M,
       author = {{Mandelker}, Nir and {Dekel}, Avishai and {Ceverino}, Daniel and {DeGraf}, Colin and {Guo}, Yicheng and {Primack}, Joel},
        title = "{Giant clumps in simulated high- z Galaxies: properties, evolution and dependence on feedback}",
      journal = {\mnras},
         year = 2017,
        month = jan,
       volume = {464},
       number = {1},
        pages = {635-665},
          doi = {10.1093/mnras/stw2358},
archivePrefix = {arXiv},
       eprint = {1512.08791},
 primaryClass = {astro-ph.GA},
       adsurl = {https://ui.adsabs.harvard.edu/abs/2017MNRAS.464..635M}
}

@ARTICLE{2022MNRAS.511.5475M,
       author = {{Marshall}, Madeline A. and {Wilkins}, Stephen and {Di Matteo}, Tiziana and {Roper}, William J. and {Vijayan}, Aswin P. and {Ni}, Yueying and {Feng}, Yu and {Croft}, Rupert A.~C.},
        title = "{The impact of dust on the sizes of galaxies in the Epoch of Reionization}",
      journal = {\mnras},
         year = 2022,
        month = apr,
       volume = {511},
       number = {4},
        pages = {5475-5491},
          doi = {10.1093/mnras/stac380},
archivePrefix = {arXiv},
       eprint = {2110.12075},
 primaryClass = {astro-ph.GA},
       adsurl = {https://ui.adsabs.harvard.edu/abs/2022MNRAS.511.5475M}
}

@ARTICLE{2016ApJ...831...78M,
       author = {{Mieda}, Etsuko and {Wright}, Shelley A. and {Larkin}, James E. and {Armus}, Lee and {Juneau}, St{\'e}phanie and {Salim}, Samir and {Murray}, Norman},
        title = "{IROCKS: Spatially Resolved Kinematics of z {\ensuremath{\sim}} 1 Star-forming Galaxies}",
      journal = {\apj},
         year = 2016,
        month = nov,
       volume = {831},
       number = {1},
          eid = {78},
        pages = {78},
          doi = {10.3847/0004-637X/831/1/78},
archivePrefix = {arXiv},
       eprint = {1608.01676},
 primaryClass = {astro-ph.GA},
       adsurl = {https://ui.adsabs.harvard.edu/abs/2016ApJ...831...78M}
}
\bibliographystyle{aasjournalv7.1}

\end{document}